\documentclass[aps,preprint,floats,epsf,epsfig,nofootinbib,letter]{revtex4}

\usepackage{graphicx}% Include figure files
\usepackage{dcolumn}% Align table columns on decimal point
\usepackage{bm}% bold math

\begin{document}
\def\be{\begin{eqnarray}}
\def\en{\end{eqnarray}}
\def\non{\nonumber}
\def\la{\langle}
\def\ra{\rangle}
\def\nc{N_c^{\rm eff}}
\def\vp{\varepsilon}
\def\a{{\cal A}}
\def\B{{\cal B}}
\def\c{{\cal C}}
\def\d{{\cal D}}
\def\e{{\cal E}}
\def\p{{\cal P}}
\def\t{{\cal T}}
\def\up{\uparrow}
\def\dw{\downarrow}
\def\vma{{_{V-A}}}
\def\vpa{{_{V+A}}}
\def\smp{{_{S-P}}}
\def\spp{{_{S+P}}}
\def\J{{J/\psi}}
\def\ov{\overline}
\def\Lqcd{{\Lambda_{\rm QCD}}}
\def\pr{{\sl Phys. Rev.}~}
\def\prl{{\sl Phys. Rev. Lett.}~}
\def\pl{{\sl Phys. Lett.}~}
\def\np{{\sl Nucl. Phys.}~}
\def\zp{{\sl Z. Phys.}~}
\def\lsim{ {\ \lower-1.2pt\vbox{\hbox{\rlap{$<$}\lower5pt\vbox{\hbox{$\sim$}
}}}\ } }
\def\gsim{ {\ \lower-1.2pt\vbox{\hbox{\rlap{$>$}\lower5pt\vbox{\hbox{$\sim$}
}}}\ } }

\font\el=cmbx10 scaled \magstep2{\obeylines\hfill March, 2010}

\vskip 1.5 cm

\centerline{\large\bf Hadronic $D$ decays involving even-parity light mesons}
\bigskip
\bigskip
\centerline{\bf Hai-Yang Cheng,$^{1,2}$ Cheng-Wei Chiang$^{3,1}$}
\medskip
\centerline{$^1$ Institute of Physics, Academia Sinica}
\centerline{Taipei, Taiwan 115, Republic of China}
\medskip
\centerline{$^2$ Physics Department, Brookhaven National Laboratory} \centerline{Upton, New York 11973}
\medskip
\centerline{$^3$ Department of Physics and Center for Mathematics } \centerline{and Theoretical Physics, National Central University}
\centerline{Chungli, Taiwan 320, Republic of China}
\medskip

\bigskip
\bigskip
\centerline{\bf Abstract}
\bigskip
\small
We study the hadronic $D$ meson decays into a pseudoscalar meson $P$ and an even-parity meson $M$, where $M$ represents a scalar meson $S$, an axial-vector meson $A$, or a tensor meson $T$.  These decays are first analyzed in the flavor-diagram approach.  Fits to the $SP$ modes with $S$ being a non-strange scalar meson show that neither the simple $q\bar q$ picture nor the $q^2\bar q^2$ scheme is favored by data.  Current measurements on the $AP$ decays are insufficient for a meaningful analysis.  Some $TP$ data are inconsistent with the others.  In certain cases, the $W$-annihilation diagrams indicated by the data are unexpectedly large.  As a comparison, we also compute their decay rates in the factorization approach using form factors extracted from the covariant light-front model.  We find that factorization works well for Cabibbo-allowed $D^+ \to SP,AP$ decays free of the weak annihilation contributions ($W$-exchange or $W$-annihilation).  For the other $SP$ and $AP$ modes, it is necessary to include weak annihilation contributions to account for the data.  However, factoriztion fails for $D\to TP$ decays for some unknown reason; the predicted rates are in general too small by at least two orders of magnitude compared to experiment.  We also examine the finite width effects of resonances.  Some decay modes which are kinematically forbidden become physically allowed due to the finite width of the resonance.  We show that the branching fraction of $D^+\to\sigma\pi^+$  extracted from three-body decays is enhanced by a factor of 2, whereas $\B(D^0\to f_2(1270)\ov K^0)$ is reduced by a factor of 4 by finite width effects.

\pagebreak

\section{Introduction}

A plethora of interesting but puzzling phenomena regarding strong interactions have been revealed in the study of charmed meson decays.  The magnitudes and phases of CKM factors associated with the dominant decay amplitudes in such processes are well-determined.  Moreover, these decays do not receive significant corrections from penguin-type loop diagrams.  Such salient features enable one to readily extract the magnitudes and relative strong phases of various flavor diagrams.  More specifically, there are four types of flavor digrams that dominate the charmed meson decays.  They are the color-allowed tree amplitude, the color-suppressed tree amplitude, the $W$-exchange diagram, and the $W$-annihilation diagram.  Previous studies \cite{Buccella95,Gronau,Rosner99,Gronau:2000ru,Chiang:2001av,Ablikim:2002ha,Chiang03,Ligeti} have shown the importance of the $W$-exchange and $W$-annihilation diagrams, presumably due to significant final-state rescattering effects, in the two-body decays of charmed mesons to two pseudoscalar ($PP$) or one vector and one pseudoscalar mesons ($VP$).  Moreover, these amplitudes are seen to have nontrivial relative strong phases, also a result of sizable final-state interactions (FSIs).  Among various sources of FSIs, the most important one is arguably the contribution from intermediate resonance states near the $D$ meson masses.

In this paper, we set to study the hardonic decays of charmed mesons into a pseudoscalar meson $P$ and an even-parity meson $M$.  Here $M$ can be a scalar meson, denoted by $S$, an axial-vector meson, denoted by $A$, or a tensor meson, denoted by $T$.  The $D \to SP$ decays have been studied previously in Refs.~\cite{Kamal,Katoch,Buccella96,Fajfer,ChengSP,ElBennich,Boito}; the $D \to AP$ decays in Refs.~\cite{Kamal:1991kg,Pham:1991az,Pham:1991ex,Kamal:1993gu,Katoch:1995bm,%
  Lipkin:2000gz,ChengAP,Sharma,Khosravi}; and the $D \to TP$ decays in Refs.~\cite{Katoch:1994zk,Munoz:1998sn,ChengTP}.  In these decays, the flavor diagram of each topology has two possibilities: one with the spectator quark in the charmed meson going to the pseudoscalar meson in the final state; and the other with the spectator quark ending up in the even-parity meson $M$.  We thus need two copies of each topological diagram to describe the decay processes.  Many of these decays have been observed in recent years through dedicated experiments and powerful Dalitz plot analysis of multi-body decays.  An extraction of the sizes and relative strong phases of these amplitudes therefore becomes possible.

One purpose of studying these decays is to check our understanding in the structures and properties of light even-parity mesons.  Another goal is to learn the FSI pattern in view of the rich resonance spectrum around the $D$ meson mass range.

This paper is organized as follows.  In Section~\ref{sec:status}, we review current experimental status of the measurements of multi-body charmed meson decays that are relevant to our analysis.  We provide the information of flavor SU(3) classification, decay constants, and form factors for the light $S$, $A$, and $T$ mesons in Section~\ref{sec:properties}.  Section~\ref{sec:flavorapp} presents the so-called flavor-diagram approach to the decays.  Each decay mode is decomposed in terms of quark diagrams characterized by their flavor topologies.  Current experimental data are used to infer the magnitude and strong phase associated with each of the amplitudes as best as we can.  Under the factorization assumption, we compute the rate of each decay mode in Section~\ref{sec:facapp}.  We also examine the finite width effects for certain decay modes in Section~\ref{sec:finitewidth}.  Some discussions on the form factor model used in this work, theoretical uncertainties, comparison with other approaches are presented in Section~\ref{sec:discussions}. A summary of our findings is given in Section~\ref{sec:conclusions}.

\section{Experimental status \label{sec:status}}

It is known that three- and four-body decays of heavy mesons provide a rich laboratory for studying the intermediate state resonances.  The Dalitz plot analysis is a very useful technique for this purpose.  We are interested in $D\to MP$  decays extracted from the three-body decays of charmed mesons.  Many results are available from ARGUS, Femilab, CLEO, FOCUS and BaBar.  The results of various experiments are summarized in Tables~\ref{tab:DataSP}-\ref{tab:TPData} where the products of $\B(D\to MP)$ and $\B(M\to P_1P_2)$ are listed.  To extract the branching fraction for $D\to MP$, we apply the narrow width approximation
 \be \label{eq:fact}
 \Gamma(D\to MP\to P_1P_2P)=\Gamma(D\to MP)\B(M\to P_1P_2) ~.
 \en
(Finite width effects in certain decays will be discussed in Section~\ref{sec:finitewidth}.)  For the branching fractions of two-body decays of even-parity mesons, we shall use
\be
 \B(f_0(980)\to\pi^+\pi^-)=0.35\pm0.08 ~, &&
 \B(a_0^+(980)\to \eta\pi)=0.845\pm 0.017 ~, \non \\
 \B(K_0^{*0}(1430)\to K^+\pi^-)={2\over 3}(0.93\pm0.10) ~, &&
 \B(K_0^{*+}(1430)\to K^+\pi^0)={1\over 3}(0.93\pm0.10) ~, \non \\
 \B(f_2(1270)\to\pi\pi) = (84.8^{+2.4}_{-1.2})\% ~, &&
 \B(f_2(1270)\to K\ov K) =(4.6\pm 0.4)\% ~,
 \non  \\
 \B(a_2(1320)\to K\ov K)=(4.9\pm 0.8)\% ~, &&
 \B(K_2^*(1430)\to K\pi)=(49.9\pm 1.2)\% ~,
 \en
and $\B(f_0(1500)\to\pi^+\pi^-)={2\over 3}(34.9\pm2.3)\%$, where we have applied the value of $\Gamma(f_0(980)\to \pi\pi) / [\Gamma(f_0(980)\to\pi\pi) + \Gamma(f_0(980)\to K\ov K)] = 0.52\pm0.12$ obtained by BaBar \cite{f0BR} and the Particle Data Group (PDG) average, $\Gamma(a_0(980)\to K\ov K)/\Gamma(a_0(980)\to \pi\eta)=0.183\pm 0.024$ \cite{PDG}.  To obtain the branching fraction of $f_0(980)\to\pi\pi$, we have assumed that $\pi\pi$ and $K\bar K$ are the dominant decay modes of $f_0(980)$.

Several remarks are in order:
\begin{enumerate}
\item There are two measurements of $D_s^+\to f_0(980)\pi^+$ from $D_s^+ \to \pi^+ \pi^+ \pi^-$ by E687 \cite{E687} and E791 \cite{E791} with the results
  \be
  \B(D_s^+\to f_0(980)\pi^+)\B(f_0\to\pi^+\pi^-)=\left\{
    \begin{array}{cl}
     (6.3\pm0.8)\times 10^{-3}
      & \quad \mbox{E791,}  \\
     (1.2\pm0.2)\times 10^{-2}
      & \quad \mbox{E687.}
    \end{array}\right.
  \en
These two data are not used by PDG for the average.  The most recent Dalitz plot analysis of $D_s^+ \to \pi^+\pi^+\pi^-$ by BaBar yields $\B(D_s^+\to (\pi^+\pi^-)_{\rm S-wave}\pi^+) = (0.92\pm0.07)\times 10^{-2}$ \cite{BaBar:Ds3pi}. Since the $S$-wave is the sum over $f_0(980),~f_0(1370)$ and $f_0(1500)$, it is clear that the E687 result is too large. Hence, we will only quote the E791 result in Table~\ref{tab:DataSP} for $D_s^+\to f_0(980)\pi^+$ with $f_0(980) \to \pi^+ \pi^-$. For an early theoretical study, see \cite{Gourdin}.

\item Many of the 3-body decays listed in Table~\ref{tab:DataSP} involve the decays $f_0(980)\to K^+K^-$ and $a_0(980)\to K^+K^-$.  Since the central values of the $f_0(980)$ and $a_0(980)$ masses are below the threshold for decay into a pair of charged kaons, the narrow width approximation Eq.~(\ref{eq:fact}) is no longer applicable because $f_0$ or $a_0$ need to be off-shell.  That is, the relation
\be
\B(D^0\to f_0(980)\pi^0; f_0\to K^+K^-)
=\B(D^0\to f_0(980)\pi^0)\B(f_0\to K^+K^-) ~,
\en
does not hold.  From Table~\ref{tab:DataSP}, we see that $\B(D^0\to f_0\pi^0; f_0\to\pi^+\pi^-)$ is smaller than $\B(D^0\to f_0\pi^0; f_0\to K^+K^-)$ by one order of magnitude.  Since $\B(f_0\to K^+K^-)$ is smaller than $\B(f_0\to\pi^+\pi^-)$ due to phase space suppression\footnote{The ratio $\B(f_0\to K^+K^-)/\B(f_0\to \pi^+\pi^-)$ was measured to be $0.69\pm0.32$ by BaBar \cite{f0BR} and  $0.25^{+0.17}_{-0.11}$ by BES \cite{BESf0}.}, the branching fraction $\B(f_0(980)\to K^+K^-)$ obtained from $\B(D^0\to f_0\pi^0; f_0\to K^+K^-)$ under the narrow width approximation will be too large by at least one order of magnitude.

\item The decay $D^+\to K^-\pi^+\pi^+$ is dominated by the $S$-wave $(K^-\pi^+)$ component which consists of $K_0^*(800)$ (or $\kappa$), $K^*_0(1430)$ and non-resonant contributions.  The PDG value is $\B(D^+\to (K^-\pi^+)_{\rm S-wave}\pi^+) = (7.62\pm0.25)\%$ \cite{PDG}.  However, PDG does not take the measurements of $D^+\to \bar K^{*0}_0(800)\pi^+$ and $\bar K^{*0}_0(1430)\pi^+$ by E791, E691 and E687 for the average.  If we take the E791 results alone \cite{E791}, we find
\be \label{eq:E791}
\B(D^+\to \bar K^{*0}_0(800)\pi^+)=(6.5\pm 1.9)\% ~, \qquad
\B(D^+\to \bar K^{*0}_0(1430)\pi^+)=(1.8\pm 0.3)\% ~.
\en
Hence, the decay $D^+\to \bar K^{*0}_0(800)\pi^+$ has the largest branching fraction among the two-body $D\to SP$ decays.

\item Since the axial-vector mesons decay into three pseudoscalar mesons via strong interactions, their resonant substructures are studied in the Dalitz plot analysis of four-body decays.  For example, information on the decay rates of $D^0\to K^-a_1^+(1260)$ and $K_1^-(1270)\pi^+$ can be extracted from the study of $D^0\to K^-\pi^+\pi^-\pi^+$.

\item Because the $\sigma$ meson is very broad in its width, of ${\cal O}(600-1000)$ MeV \cite{PDG}, the use of the narrow width approximation is not justified and it becomes necessary to take into account the finite width effect of $\sigma$.  We will examine the finite width effect for the decay $D\to\sigma\pi$ in Section~\ref{sec:finitewidth}.

\end{enumerate}

{\squeezetable
\begin{table}[t]
\caption{Experimental branching fractions of various $D\to SP$ decays.  For simplicity and convenience, we have dropped the mass identification for
$f_0(980)$, $a_0(980)$ and $K^*_0(1430)$.  Data are taken from Ref.~\cite{PDG} unless specified otherwise.}
\label{tab:DataSP}
\begin{ruledtabular}
\begin{tabular}{l l l}
$\B(D\to SP)\times \B(S\to P_1P_2)$  & $\B(D\to SP)$ \\ \hline
 $\B(D^+\to f_0\pi^+)\B(f_0\to\pi^+\pi^-)=(1.55\pm 0.33)\times 10^{-4}$ &
 $\B(D^+\to f_0\pi^+)=({4.5}\pm 1.4)\times 10^{-4}$ \\
 $\B(D^+\to f_0(1370)\pi^+)\B(f_0(1370)\to\pi^+\pi^-)=(8\pm4)\times 10^{-5}$ & $$ \\
 $\B(D^+\to f_0(1500)\pi^+)\B(f_0(1500)\to\pi^+\pi^-)=(1.1\pm0.4)\times 10^{-4}$ & $\B(D^+\to f_0(1500)\pi^+)=(4.7\pm1.7)\times 10^{\bf -4}$ \\
 $\B(D^+\to f_0(1710)\pi^+)\B(f_0(1710)\to\pi^+\pi^-)<5\times 10^{-5}$ & $$ \\
 $\B(D^+\to f_0K^+)\B (f_0\to \pi^+\pi^-)=(5.6\pm 3.4)\times 10^{-5}$ &
 $\B(D^+\to f_0K^+)=(1.6\pm 1.0)\times 10^{-4}$ \\
 $\B(D^+\to\sigma\pi^+)\B(\sigma\to\pi^+\pi^-)=(1.37\pm0.12)\times 10^{-3}$ &
 $\B(D^+\to\sigma\pi^+)=({2.1\pm0.2})\times 10^{-3}$ \\
 $\B(D^+\to \bar\kappa^0 K^+)\B(\bar \kappa^0\to K^-\pi^+)=(6.8^{+3.5}_{-2.1})\times 10^{-4}$ &
 $\B(D^+\to \bar\kappa^0 K^+)=({1.0^{+0.5}_{-0.3}})\times 10^{-3}$ \\
%
% $\B(D^+\to\ov K_0^{*0}\pi^+)\B(\ov K_0^{*0}\to K^-\pi^+)=(2.41\pm 0.24)\%$  &
% $\B(D^+\to\ov K_0^{*0}\pi^+)=(3.89\pm0.39)\%$ \\
%
 $\B(D^+\to\ov K_0^{*0}K^+;\ov K_0^{*0}\to K^-\pi^+)=(1.83\pm 0.35)\times 10^{-3}$  &
 $$ \\
 $\B(D^+\to (K^-\pi^+)_{\rm S-wave}\pi^+)=(7.62\pm0.25)\%$ &\\
  \hline
 $\B(D^0\to f_0\ov K^0)\B(f_0\to \pi^+\pi^-)=(2.76^{+0.60}_{-0.44})\times 10^{-3}$  &
 $\B(D^0\to f_0\ov K^0)=({ 8.0}^{+2.5}_{-2.2})\times 10^{-3}$ \\
 $\B(D^0\to f_0(1370)\ov K^0)\B(f_0(1370)\to \pi^+\pi^-)
 =(5.0\pm1.2)\times 10^{-3}$  & \\
 $\B(D^0\to f_0\pi^0)\B(f_0\to \pi^+\pi^-)=(3.6\pm0.8)\times 10^{-5}$  &
 $\B(D^0\to f_0\pi^0)=(1.0\pm 0.3)\times 10^{-4}$ \\
 $\B(D^0\to f_0\pi^0; f_0\to K^+K^-)=(3.5\pm0.6)\times 10^{-4}$  &
  \\
 $\B(D^0\to f_0(1370)\pi^0)\B(f_0\to \pi^+\pi^-)=(5.3\pm2.1)\times 10^{-5}$  &
  \\
 $\B(D^0\to f_0(1500)\pi^0)\B(f_0\to \pi^+\pi^-)=(5.6\pm1.6)\times 10^{-5}$  &
 $\B(D^0\to f_0(1500)\pi^0)=(2.4\pm0.7)\times 10^{-4}$ \\
 $\B(D^0\to f_0(1710)\pi^0)\B(f_0\to \pi^+\pi^-)=(4.5\pm1.5)\times 10^{-5}$  &
  \\
 $\B(D^0\to f_0\ov K^0;f_0\to K^+K^-)<2.0\times 10^{-4}$  &  \\
 $\B(D^0\to f_0(1370)\ov K^0)\B(f_0\to K^+K^-)=(3.6\pm2.2)\times 10^{-4}$  &  \\
 $\B(D^0\to a_0^+K^-;a_0^+\to K^+\ov K^0)=(1.24\pm 0.36)\times 10^{-3}$ & \\
 $\B(D^0\to a_0^+K^-;a_0^+\to K^+\ov K^0)=(1.5\pm 0.1)\times 10^{-3}$ \footnotemark[1] & \\
 $\B(D^0\to a_0^+(1450)K^-;a_0^+\to K^+\ov K^0)=(2.1\pm 0.1)\times 10^{-3}$ \footnotemark[1] & \\
 $\B(D^0\to a_0^-K^+;a_0^-\to K^-\ov K^0)<2.4\times 10^{-4}$ &  \\
 $\B(D^0\to a_0^0\ov K^0;a_0^0\to K^+K^-)=(6.2\pm 0.8)\times 10^{-3}$ &  \\
 $\B(D^0\to a_0^0\ov K^0;a_0^0\to K^+K^-)=(5.3\pm 0.4)\times 10^{-3}$ \footnotemark[1]  &  \\
 $\B(D^0\to a_0^0\ov K^0)\B(a_0^0\to
 \eta\pi^0)=(1.34\pm0.42)\times 10^{-2}$ & $\B(D^0\to a_0^0\ov K^0)=({ 1.6\pm0.5})\%$ \\
 $\B(D^0\to \sigma\pi^0)\B(\sigma\to \pi^+\pi^-)=(1.18\pm0.21)\times 10^{-4}$  &
 $\B(D^0\to \sigma\pi^0)=({1.8\pm0.3})\times 10^{-4}$ \\
 $\B(D^0\to K_0^{*-}\pi^+)\B(K_0^{*-}\to \ov K^0\pi^-)=(4.90^{+0.80}_{-0.64})\times 10^{-3}$ &
 $\B(D^0\to K_0^{*-}\pi^+)=({ 7.9^{+1.5}_{-1.3}})\times 10^{-3}$ \\
 $\B(D^0\to K_0^{*-}\pi^+)\B(K_0^{*-}\to K^-\pi^0)=(4.6\pm 2.1)\times 10^{-3}$ &
 $\B(D^0\to K_0^{*-}\pi^+)=(1.5\pm0.7)\%$ \\
 $\B(D^0\to \ov K_0^{*0}\pi^0)\B(\ov K_0^{*0}\to K^-\pi^+)=
 (5.7^{+5.0}_{-1.5})\times 10^{-3}$ &
 $\B(D^0\to \ov K_0^{*0}\pi^0)=(9.2^{+8.1}_{ -2.6})\times 10^{-3}$  \\
  \hline
 $\B(D_s^+\to (\pi^+\pi^-)_{\rm S-wave}\pi^+)=(0.92\pm 0.07)\times 10^{-2}$ \footnotemark[2] &  \\
 $\B(D_s^+\to f_0\pi^+)\B(f_0\to\pi^+\pi^-)=(6.3\pm 0.8)\times 10^{-3}$  \footnotemark[3] & $\B(D_s^+\to f_0\pi^+)=(1.8\pm{ 0.5})\%$ \\
 $\B(D_s^+\to f_0\pi^+;f_0\to K^+K^-)=(6.0\pm 2.4)\times 10^{-3}$ &  \\
 $\B(D_s^+\to f_0\pi^+;f_0\to K^+K^-)=(1.55\pm 0.13)\times 10^{-2}$ \footnotemark[4] &  \\
 $\B(D_s^+\to f_0(1370)\pi^+)\B(f_0\to K^+K^-)=(2.37\pm 0.35)\times 10^{-2}$ \footnotemark[4] &  \\
 $\B(D_s^+\to f_0(1710)\pi^+)\B(f_0\to K^+K^-)=(1.87\pm 0.29)\times 10^{-2}$ \footnotemark[4] &  \\
 $\B(D_s^+\to \ov K_0^{*0}K^+)\B(\ov K_0^{*0}\to K^-\pi^+)=(5.1\pm2.5)\times 10^{-3}$ &
 $\B(D_s^+\to \ov K_0^{*0}K^+)=({ 8.2\pm4.1})\times 10^{-3}$  \\
 $\B(D_s^+\to \ov K_0^{*0}K^+)\B(\ov K_0^{*0}\to K^-\pi^+)=(2.15\pm0.40)\times 10^{-3}$ \footnotemark[4] &
 $\B(D_s^+\to \ov K_0^{*0}K^+)=(3.5\pm0.7)\times 10^{-3}$  \\
 $\B(D_s^+\to K_0^{*0}\pi^+)\B(K_0^{*0}\to K^+\pi^-)=(5\pm4)\times 10^{-4}$  &
 $\B(D_s^+\to K_0^{*0}\pi^+)=(8.1\pm6.5)\times 10^{-4}$  \\
\end{tabular}
\footnotetext[1]{From BaBar \cite{BaBar:Da0K}.}
\footnotetext[2]{From BaBar \cite{BaBar:Ds3pi}.}
\footnotetext[3]{From E791 \cite{E791}.}
\footnotetext[4]{From CLEO \cite{CLEO:Dsf0pi}.}
\end{ruledtabular}
\end{table}
}

\begin{table}[t]
\caption{Experimental branching fractions of $D\to AP$ decays taken from Ref.~\cite{PDG}.}
\label{tab:}
\begin{ruledtabular}
\begin{tabular}{ l  c  | l  c  }
Decay~~~~~
    & ~~~Experiment~~~~~~~~
    &  Decay~~~~~
& ~~~ Experiment~~~
 \\
    \hline
$D^+\to \bar K_1^0(1270)\pi^+$
    & $<7\times 10^{-3}$
& $D^+\to \bar K^0 a_1^+(1260)$
    & $(7.0\pm1.2)\%$
    \\
$D^+\to \bar K_1^0(1400)\pi^+$
    & $(3.8\pm1.3)\%$
& $D^0\to K^- a_1^+(1260)$
    & $(7.9\pm1.1)\%$
    \\
$D^0\to K_1^-(1270)\pi^+$
    & $(1.14\pm0.32)\%$
& $D^0\to \bar K^0 a_1^0(1260)$
    & $<1.9\%$
    \\
$D^0\to K_1^-(1400)\pi^+$
    & $<1.2\%$
& $D^0\to \pi^- a_1^+(1260)$
    & $(8.98\pm0.62)\times 10^{-3}$
    \\
$D^0\to \bar K_1^0(1400)\pi^0$
    & $<3.7\%$
& &
\end{tabular}
\begin{tabular}{ l l}
$\B(D^0\to K_1^\pm(1270)K^\mp)\B(K_1^\pm(1270)\to K^\pm\pi^+\pi^-)=(8.1\pm1.8)\times 10^{-4}$ \\
$\B(D^0\to K_1^\pm(1400)K^\mp; K_1^\pm(1400)\to K^\pm\pi^+\pi^-)=(5.4\pm1.2)\times 10^{-4}$ \\
\end{tabular}
\end{ruledtabular}
\end{table}

\begin{table}[pth]
\caption{Experimental branching fractions of various $D\to TP$ decays.  For
simplicity and convenience, we have dropped the mass
identification for $f_2(1270)$, $a_2(1320)$ and $K^*_2(1430)$.  Data are taken from Ref.~\cite{PDG} unless specified otherwise.}
\label{tab:TPData}
\begin{ruledtabular}
\begin{tabular}{ l l l }
$\B(D\to TP)\times \B(T\to P_1P_2)$  & $\B(D\to TP)$ \\
 \hline
 $\B(D^+\to f_2\pi^+)\B(f_2\to\pi^+\pi^-)=(5.0\pm 0.9)\times 10^{-4}$ &
 $\B(D^+\to f_2\pi^+)=({ 8.8\pm 1.6})\times 10^{-4}$ \\
 $\B(D^+\to \ov K^{*0}_2\pi^+)\B(\ov K_2^{*0}\to K^-\pi^+)=(2.1\pm 0.4)\times 10^{-4}$ &
 $\B(D^+\to \ov K^{*0}_2\pi^+)=(6.3\pm 1.2)\times 10^{-4}$ \\
 $\B(D^+\to K^{*0}_2\pi^+)\B(K_2^{*0}\to K^+\pi^-)=(5.0\pm 3.4)\times 10^{-5}$ &
 $\B(D^+\to K^{*0}_2\pi^+)=(1.5\pm 1.0)\times 10^{-4}$ \\
 $\B(D^+\to \ov K^{*0}_2 K^+; \ov K_2^{*0}\to K^-\pi^+)=(1.7^{+1.2}_{-0.8})\times 10^{-4}$ &
 $$ \\
 $$ &
 $\B(D^+\to a_2^+\ov K^0)<3.0\times 10^{-3}$ \\
 \hline
 $\B(D^0\to f_2\pi^0)\B(f_2\to\pi^+\pi^-)=(1.91\pm 0.20)\times 10^{-4}$ &
 $\B(D^0\to f_2\pi^0)=(3.4\pm 0.4)\times 10^{-4}$ \\
 $\B(D^0\to f_2\ov K^0)\B(f_2\to\pi^+\pi^-)=(2.8^{+2.0}_{-1.2})\times 10^{-4}$ &
 $\B(D^0\to f_2\ov K^0)=(5.0^{+3.5}_{-2.1})\times 10^{-4}$ \\
 $\B(D^0\to K^{*-}_2\pi^+)\B(K_2^{*-}\to \ov K^0\pi^-)=(7.0^{+4.0}_{-2.2})\times 10^{-4}$ &
 $\B(D^0\to K^{*-}_2\pi^+)=(2.1^{+1.2}_{ -0.7})\times 10^{-3}$ \\
 $$ &
 $\B(D^0\to a_2^+K^-)<2\times 10^{ -3}$ \\
 \hline
 $\B(D_s^+\to f_2\pi^+)\B(f_2\to\pi^+\pi^-)=(1.1\pm 0.2)\times 10^{-3}$ \footnotemark[1] &
 $\B(D_s^+\to f_2\pi^+)=({ 1.9}\pm 0.4)\times 10^{-3}$ \\
 $\B(D_s^+\to K_2^{*0}\pi^+)\B(K_2^{*0}\to K^+\pi^-)=(5\pm4)\times 10^{-4}$ &
 $\B(D_s^+\to K_2^{*0}\pi^+)=(1.5\pm1.2)\times 10^{-3}$  \\
\end{tabular}
\footnotetext[1]{From BaBar \cite{BaBar:Ds3pi}.}
\end{ruledtabular}
\end{table}

\section{Physical properties of even-parity mesons \label{sec:properties}}

\subsection{Scalar mesons \label{sec:S}}

It is known that the underlying structure of scalar mesons is not well established theoretically (for a review, see {\it e.g.} Refs.~\cite{Amsler,Close}).  Many scalar mesons with masses lower than 2 GeV have been observed, and they can be classified into two nonets: one nonet with masses below or close to 1 GeV, namely, the isoscalars $f_0(600)$ (or $\sigma$), $f_0(980)$, the isodoublet $K_0^*(800)$ (or $\kappa$) and the isovector $a_0(980)$; and the other nonet with masses above 1 GeV, namely, $f_0(1370)$, $a_0(1450)$, $K^*_0(1430)$ and $f_0(1500)/f_0(1710)$\footnote{Since not all three isosinglet scalars $f_0(1710)$, $f_0(1500)$, $f_0(1370)$ can be accommodated in the $q\bar q$ nonet picture, it is widely believed that one of them should be primarily a scalar glueball.}.  If the scalar meson states below or near 1 GeV are identified as a conventional low-lying $0^+$ $q\bar q$ nonet, then the nonet states above 1 GeV could be excited $q\bar q$ states.

In the naive quark model, the flavor wave functions of the light scalars read
 \be
 && \sigma={1\over \sqrt{2}}(u\bar u+d\bar d) ~, \qquad\qquad
 f_0= s\bar s ~, \non \\
 && a_0^0={1\over\sqrt{2}}(u\bar u-d\bar d) ~, \qquad\qquad a_0^+=u\bar d ~,
 \qquad a_0^-=d\bar u ~,  \\
 && \kappa^{+}=u\bar s ~, \qquad \kappa^{0}= d\bar s ~, \qquad~
 \bar \kappa^{0}=s\bar d ~,\qquad~ \kappa^{-}=s\bar u ~, \non
 \en
where the ideal mixing for $f_0$ and $\sigma$ is assumed as $f_0(980)$ is the heaviest one and $\sigma$ the lightest one in the light scalar nonet.  However, this simple picture encounters several serious problems:
(i) It is impossible to understand the mass degeneracy betweem $f_0(980)$ and $a_0(980)$.  A related question is why $a_0$ is heavier than $\kappa$ if it does not contain a strange quark?  This is the so-called ``inverted spectrum problem.''
(ii) The $P$-wave $0^+$ meson has a unit of orbital angular momentum which costs energy around 500 MeV.  Hence, it should have a higher mass above rather than below 1 GeV.
(iii) It is hard to explain why $\sigma$ and $\kappa$ are much broader than $f_0(980)$ and $a_0(980)$.
(iv) The $\gamma\gamma$ widths of $a_0(980)$ and $f_0(980)$ are much smaller than naively expected for a $q\bar{q}$ state~\cite{bar85}.
(v) The radiative decay $\phi\to a_0(980)\gamma$, which cannot proceed if $a_0(980)$ is a pure $q\bar q$ state, can be nicely described in the kaon loop mechanism \cite{Schechter06}.  This suggests a considerable admixture of the  $K\bar K$ component.

It turns out that these difficulties can be readily resolved in the tetraquark scenario where the four-quark flavor wave functions of light scalar mesons are symbolically given by \cite{Jaffe}
 \be \label{4quarkw.f.}
 && \sigma=u\bar u d\bar d ~, \qquad\qquad\qquad
 f_0=s\bar s(u\bar u+d\bar d)/\sqrt{2} ~,  \non \\
 && a_0^0={1\over\sqrt{2}}(u\bar u-d\bar d)s\bar s ~,
 \qquad a_0^+=u\bar ds\bar s ~,
 \qquad a_0^-=d\bar us\bar s ~, \non \\
 && \kappa^+=u\bar sd\bar d ~, \qquad \kappa^0=d\bar su\bar u ~,
 \qquad \bar \kappa^0=s\bar du\bar u ~,
 \qquad \kappa^-=s\bar ud\bar d ~.
 \en
The four quarks $q^2\bar q^2$ can form an $S$-wave (not $P$-wave!) $0^+$ meson without introducing a unit of orbital angular momentum.  Moreover, color and spin dependent interactions favor a flavor nonet configuration with attraction between the $qq$ and $\bar q\bar q$ pairs.  Therefore, the $0^+$ $q^2\bar q^2$ nonet has a mass near or below 1 GeV.  This four-quark description explains naturally the inverted mass spectrum of the light nonet, especially the mass degeneracy of $f_0(980)$ and $a_0(980)$.  The fall-apart strong decays $\sigma\to\pi\pi$, $\kappa\to K\pi$ and $f_0,a_0\to K\ov K$ are OZI super-allowed without the need of any gluon exchange.  This explains the broad widths of $\sigma$ and $\kappa$, while $f_0(980)$ and $a_0(980)$ are narrow because of the suppressed phase space for their decays to the kaon pairs.  The decays of $f_0(980)$ and $a_0(980)$ are dominated by $f_0(980)\to\pi\pi$ and $a_0(980)\to\eta\pi$, respectively.  Lattice calculations have confirmed that $a_0(1450)$ and $K_0^*(1430)$ are $q\bar q$ mesons, and suggested that $\sigma$ and $\kappa$ are tetraquark mesonia \cite{Prelovsek,Mathur}.  Since exotic 4-quark states have not been seen experimentally, this may imply the structure of diquark-antidiquark bound states for the light scalar mesons (for a review, see Ref.~\cite{Jaffe:exotic}).

In the 2-quark picture with ideal mixing, $f_0(980)$ is purely an $s\bar s$ state.  This is supported by the data of $D_s^+\to f_0\pi^+$ and $\phi\to f_0\gamma$, implying the copious $f_0(980)$ production via its $s\bar s$ component.  However, there also exists some experimental evidence indicating that $f_0(980)$ is not a pure $s\bar s$ state.  First, the observation of $\Gamma(J/\psi\to f_0\omega)\approx {1\over 2}\Gamma(J/\psi\to f_0\phi)$ \cite{PDG} clearly shows the existence of the non-strange and strange quark contents in $f_0(980)$.  Second, the facts that $f_0(980)$ and $a_0(980)$ have similar widths and that the $f_0$ width is dominated by $\pi\pi$ also suggest the composition of $u\bar u$ and $d\bar d$ pairs in $f_0(980)$; that is, $f_0(980)\to\pi\pi$ should not be OZI suppressed relative to $a_0(980)\to\pi\eta$.  Therefore, isoscalars $\sigma(600)$ and $f_0(980)$ should have a mixing
 \be \label{eq:mixing}
 |f_0(980)\ra = |s\bar s\ra\cos\theta+|n\bar n\ra\sin\theta ~,
 \qquad |\sigma(600)\ra = -|s\bar s\ra\sin\theta+|n\bar n\ra\cos\theta ~,
 \en
with $n\bar n\equiv (\bar uu+\bar dd)/\sqrt{2}$.  Experimental implications for the $f_0$-$\sigma$ mixing angle have been discussed in detail in Ref.~\cite{ChengSP}: the mixing angle lies in the ranges of $25^\circ<\theta<40^\circ$ and $140^\circ<\theta< 165^\circ$\footnote{Recently CLEO has measured the semileptonic decay $D_s^+\to f_0(980)e^+\nu_e$ with the result $\B(D_s^+\to f_0(980)e^+\nu_e)\B(f_0\to\pi^+\pi^-)=(0.20\pm0.03\pm0.01)\%$ \cite{CLEO:semiDs}.  Using the value $\B(f_0\to\pi^+\pi^-)=(50^{+7}_{-9})\%$ inferred from the BES measurement \cite{BESf0} and the QCD sum rule prediction $\B(D_s^+\to f_0 e^+\nu)=\cos^2\theta\times(0.41)\%$ \cite{Aliev}, CLEO then extracted the mixing angle to be $\cos^2\theta=0.98^{+0.02}_{-0.21}$.  However, this is subject to two major uncertainties.  First, the branching fraction of $f_0(980)\to\pi^+\pi^-$ has not been measured directly.  For $\B(f_0(980)\to\pi^+\pi^-)\approx 0.35$ as used in this work, $\B(D_s^+\to f_0(980)e^+\nu_e)$ and $\cos^2\theta$ would be enhanced by a factor of $1.4$. Second, the theoretical prediction of this semileptonic decay is model dependent as it depends on the form factor of the $D_s^+\to f_0^s$ transition with $f_0^s=s\bar s$.}.

Likewise, in the four-quark scenario for light scalar mesons, one can also define a similar $f_0$-$\sigma$ mixing angle
  \be
 |f_0(980)\ra =|n\bar ns\bar s\ra\cos\phi
 +|u\bar u d\bar d\ra\sin\phi ~, \qquad
 |\sigma(600)\ra = -|n\bar ns \bar s\ra\sin\phi+|u\bar u d\bar d\ra\cos\phi ~.
 \en
It has been shown that $\phi=174.6^\circ$ \cite{Maiani}.

In principle, the 2-quark and 4-quark descriptions of the light scalars can be discriminated in the semileptonic charm decays.  For example, the ratio
\be
R={\B(D^+\to f_0\ell^+\nu)+\B(D^+ \to \sigma\ell^+\nu)
\over \B(D^+\to a_0^0\ell^+\nu)}
\en
is equal to 1 in the 2-quark scenario and 3 in the 4-quark model under the flavor SU(3) symmetry \cite{CDLu}.  In reality, the light scalar mesons may have both 2-quark and 4-quark components.  Indeed, a real hadron in the QCD language should be described by a set of Fock states each of which has the same quantum number as the hadron.  For example,
\begin{eqnarray}\label{eq:fockexpansion}
|a^+(980)\rangle &=& \psi_{u\bar d}^{a_0} |u\bar d\rangle +
\psi_{u\bar dg}^{a_0} |u\bar d g\ra + \psi_{u\bar d s\bar s}^{a_0}
|u\bar d s \bar s\rangle+ \dots\,.
\end{eqnarray}
In the tetraquark model, $\psi_{u\bar d s\bar s}^{a_0} \gg \psi_{u\bar d}^{a_0}$, while it is the other way around in the 2-quark model.

The decay constant of the scalar meson is defined as\footnote{For pseudoscalar mesons, the decay constant is defined as $\la P(p)|\bar q_2\gamma_\mu\gamma_5 q_1|0\ra=-if_Pp_\mu$.}
 \be \label{eq:Sdecayc}
 \la S(p)|\bar q_2\gamma_\mu q_1|0\ra=f_S p_\mu ~,
 \qquad \la S|\bar q_2q_1|0\ra=m_S\bar f_S ~.
 \en
The neutral scalar mesons $\sigma$, $f_0$ and $a_0^0$ cannot be produced via the vector current owing to charge conjugation invariance or conservation of vector current:
 \be
 f_{\sigma}=f_{f_0}=f_{a_0^0}=0 ~.
 \en
Applying the equation of motion to Eq.~(\ref{eq:Sdecayc}) yields
 \be \label{eq:EOM}
 \mu_Sf_S=\bar f_S ~, \qquad\quad{\rm with}~~\mu_S={m_S\over
 m_2(\mu)-m_1(\mu)} ~,
 \en
where $m_{2}$ and $m_{1}$ are the running current quark masses.  Therefore, the vector decay constant of the scalar meson $f_S$ vanishes in the SU(3) or isospin limit. The vector decay constants of $K^*_0(1430)$ and the charged $a_0(980)$ are non-vanishing, but they are suppressed due to the small mass difference between the constituent $s$ and $u$ quarks and between $d$ and $u$ quarks, respectively.     The scalar decay constants $\bar f_S$ have been computed in Ref.~\cite{CCY} within the framework of QCD sum rules.  From Eq.~(\ref{eq:EOM}) we obtain $f_{a_0(980)^\pm}=1.0\,{\rm MeV}$, $f_{a_0(1450)^\pm}=5.3\,{\rm MeV}$, and $f_{K^*_0(1430)}=35.9\,{\rm MeV}$.  In short, the vector decay constants of scalar mesons are either zero or small.

Form factors for $D\to P,S$ transitions are defined by \cite{BSW}
 \be \label{DSm.e.}
 \la P(p')|V_\mu|D(p)\ra &=& \left(P_\mu-{m_D^2-m_P^2\over q^2}\,q_ \mu\right)
F_1^{DP}(q^2)+{m_D^2-m_P^2\over q^2}q_\mu\,F_0^{DP}(q^2) ~, \non \\
\la S(p')|A_\mu|D(p)\ra &=& -i\Bigg[\left(P_\mu-{m_D^2-m_S^2\over
q^2}\,q_ \mu\right) F_1^{DS}(q^2)   +{m_D^2-m_S^2\over
q^2}q_\mu\,F_0^{DS}(q^2)\Bigg] ~,
 \en
where $P_\mu=(p+p')_\mu$, $q_\mu=(p-p')_\mu$.  As shown in Ref.~\cite{CCH}, a factor of $(-i)$ is needed in the $D\to S$ transition in order for the $D\to S$ form factors to be positive.  This can also be checked from heavy quark symmetry consideration \cite{CCH}.

Throughout this paper, we use the 3-parameter parametrization
 \be \label{eq:FFpara}
 F(q^2)=\,{F(0)\over 1-a(q^2/m_D^2)+b(q^2/m_D^2)^2}
 \en
for $D\to M$ transitions.  The parameters $F(0)$, $a$ and $b$ for $D \to S$ transitions calculated in the covariant light-front (CLF) quark model are exhibited in Table~\ref{tab:FFDtoS}.

\begin{table}[t]
\caption{Parameters in the form factors of $D, D_s\to f_0(980), K_0^*(1430)$ transitions in the parametrization of Eq.~(\ref{eq:FFpara}), as obtained by fitting to the covariant light-front model \cite{CCH}. The numbers in parentheses are the form factors at $q^2=0$ obtained using the ISGW2 model \cite{ISGW2}.}
 \label{tab:FFDtoS}
\begin{ruledtabular}
\begin{tabular}{| l c c c || l c c c |}
~~~$F$~~~~~
    & $F(0)$~~~~~
    & $a$~~~~~
    & $b$~~~~~~
& ~~~ $F$~~~~~
    & $F(0)$~~~~~
    & $a$~~~~~
    & $b$~~~~~~
 \\
    \hline
$F^{Df_{0q}}_0$
    & $0.49~(0.13)$
    & $0.07$
    &  {\bf $-0.03$} &
$F^{D_sf_{0s}}_0$
    & $0.46~(0.23)$
    & $-0.29$
    & $0.07$
    \\
$F^{DK_0^*(1430)}_0$
    & $0.48~(0.08)$
    & $-0.11$
    & $0.02$ &
$F^{D_sK_0^*(1430)}_0$
    & $0.51~(0.14)$
    & $0.07$
    & $0.02$
    \\
\end{tabular}
\end{ruledtabular}
\end{table}

\subsection{Axial-vector mesons \label{sec:A}}

In the quark model, two nonets of $J^P=1^+$ axial-vector mesons are expected as the orbital excitation of the $q\bar q$ system.  In terms of the spectroscopic notation $^{2S+1}L_J$, there are two types of $P$-wave axial-vector mesons, namely, $^3P_1$ and $^1P_1$.  These two nonets have distinctive $C$ quantum numbers for the corresponding neutral mesons, $C=+$ and $C=-$, respectively.  Experimentally, the $J^{PC}=1^{++}$ nonet consists of $a_1(1260)$, $f_1(1285)$, $f_1(1420)$ and $K_{1A}$, while the $1^{+-}$ nonet contains $b_1(1235)$, $h_1(1170)$, $h_1(1380)$ and $K_{1B}$.  The physical mass eigenstates $K_1(1270)$ and $K_1(1400)$ are mixtures of the $K_{1A}$ and $K_{1B}$ states (we follow PDG \cite{PDG} to denote the $^3P_1$ and $^1P_1$ states of $K_1$ by $K_{1A}$ and $K_{1B}$, respectively),
 \be \label{eq:K1mixing}
 K_1(1270)=K_{1A} \sin\theta_{K_1}+K_{1B}\cos\theta_{K_1} ~,
 \nonumber\\
 K_1(1400)=K_{1A} \cos\theta_{K_1}-K_{1B}\sin\theta_{K_1} ~.
 \en
Since these states are not charge conjugation eigenstates, consequently, mixing is not prohibited.  Indeed, the mixing is governed by the mass difference between the strange and non-strange light quarks.  There exist several estimations on the mixing angle $\theta_{K_1}$ in the literature.  From the early experimental information on masses and the partial rates of $K_1(1270)$ and $K_1(1400)$, Suzuki found two possible solutions, each with a two-fold ambiguity, $|\theta_{K_1}|\approx 33^\circ$ and $57^\circ$ \cite{Suzuki}.  A similar constraint $35^\circ\lsim |\theta_{K_1}|\lsim 55^\circ$ was obtained in Ref.~\cite{Goldman} based solely on two parameters: the mass difference between the $a_1$ and $b_1$ mesons and the ratio of the constituent quark masses.  An analysis of $\tau\to K_1(1270)\nu_\tau$ and $K_1(1400)\nu_\tau$ decays also yielded the mixing angle to be $\approx 37^\circ$ or $58^\circ$ with a two-fold ambiguity \cite{Cheng:DAP}.  Most of these estimations were obtained by assuming a vanishing $f_{K_{1B}}$.  With the help of analytical expressions of $f_{K_{1A,1B}}$ obtained in the CLF quark model \cite{CCH}, two solutions for the $K_1(1270)$-$K_1(1400)$ mixing angle, $50.8^\circ$ and $-44.8^\circ$, have been found in Ref.~\cite{Cheng:BAgamma}.  However, the second solution has been ruled out by the measurements of $B\to K_1(1270)\gamma$ and $B\to K_1(1400)\gamma$ \cite{Cheng:BAgamma}. Therefore, we shall use $\theta_{K_1}=50.8^\circ$ in the ensuing discussions.

For the decay constants and the form factors of the axial vector mesons, we shall follow Ref.~\cite{CCH} to define them as%
\footnote{The relative signs of the decay constants, form factors and mixing angles of the axial-vector mesons were often very confusing in the literature.  As stressed in Ref.~\cite{CK0708}, the sign of the mixing angle $\theta_{K_1}$ is intimately related to the relative sign of the $K_{1A}$ and $K_{1B}$ states.  In the CLF quark model \cite{CCH} and in pQCD \cite{Lu:BtoP}, the decay constants of $K_{1A}$ and $K_{1B}$ are of opposite signs, while the $D(B)\to K_{1A}$ and $D(B)\to K_{1B}$ form factors are of the same sign.  The mixing angle $\theta_{K_1}$ is positive.  It is the other way around in the approaches of QCD sum rules \cite{Yang:BtoP} and the ISGW model \cite{ISGW,ISGW2}: the decay constants of $K_{1A}$ and $K_{1B}$ have the same sign, while the $D(B)\to K_{1A}$ and $D(B)\to K_{1B}$ form factors are opposite in sign.  These two conventions are related via a redefinition of the $K_{1A}$ or $K_{1B}$ state, {\it i.e.}, $K_{1A}\to -K_{1A}$ or $K_{1B}\to -K_{1B}$.}
 \be \label{eq:deccCCH}
 \la A(p,\varepsilon)|A_\mu|0\ra &=& f_{A}m_{A}\epsilon_{\mu}^{*} ~, \non \\
 \langle{A}(p, \varepsilon)|A_\mu|{D} (p_D)\rangle
 &=& \frac{2}{m_D - m_{A}} \epsilon_{\mu\nu\alpha\beta} \epsilon^{*\nu}
 p_D^\alpha p^{\beta} A^{DA}(q^2) ~,
\nonumber \\
  \langle A (p,\varepsilon)|V_\mu|{ D}(p_D)\rangle
 &=& -i \Bigg\{ (m_D - m_{A}) \epsilon^{*}_{\mu} V_1^{DA}(q^2)
 - (\epsilon^{*}\cdot p_D)
(p_D + p)_\mu \frac{V_2^{DA}(q^2)}{m_D - m_{A}} \non \\
&&  -2 m_{A} \frac{\epsilon^{*}\cdot p_D}{q^2} q_\mu
\left[V_3^{DA}(q^2) - V_0^{DA}(q^2)\right]\Bigg\} ~.
 \en
Because of the charge conjunction invariance, the decay constant of the $^1P_1$ non-strange neutral meson such as $b_1^0(1235)$ must be zero.  In the isospin limit, the decay constant of the charged $b_1$ vanishes due to the fact that $b_1$ has an even $G$-parity and that the relevant weak axial-vector current is odd under $G$ transformation.  As for the strange axial vector mesons, it is known that the decay constant of the $^1P_1$ meson vanishes in the SU(3) limit \cite{Suzuki}.

In the following, we shall take $f_{a_1}=238\pm10$ MeV obtained using the QCD sum rule method \cite{YangNP}, similar to the $\rho$ meson, $f_\rho\approx$ 216 MeV.  This means that the $a_1(1260)$ meson can be regarded as the scalar partner of the $\rho$ meson, as it should be.  In the CLF quark model \cite{CCH}, if we increase the constituent $d$ quark mass by an amount of $5\pm2$ MeV relative to the $u$ quark mass, we find $f_{b_1}=0.6\pm0.2$ MeV for the charged $b_1$ which is very small.  Using the experimental results $\B(\tau\to K_1(1270)\nu_\tau) = (4.7\pm1.1)\times 10^{-3}$ and $\Gamma(\tau\to K_1(1270)\nu_\tau) / [\Gamma(\tau\to K_1(1270)\nu_\tau)+\Gamma(\tau\to K_1(1400)\nu_\tau)] = 0.69\pm0.15$ \cite{PDG}, we obtain
 \be
  |f_{K_1(1270)}|=169.5^{+18.8}_{-21.2}~{\rm MeV}~, \quad
  |f_{K_1(1400)}|=139.2^{+41.3}_{-45.6}~{\rm MeV}~.
 \label{eq:fK1abs}
 \en
In the CLF quark model the signs of the decay constants $f_{K_{1A}}$ and $f_{K_{1B}}$ are fixed: $f_{K_{1A}}=-212$ MeV and $f_{K_{1B}}=12$ MeV \cite{Cheng:BAgamma}. This together with the mixing angle $\theta_{K1}=50.8^\circ$ also fixes the signs of $f_{K_1}$ to be
 \be \label{eq:fK1}
 f_{K_1(1270)}=-170~{\rm MeV} ~, \quad
 f_{K_1(1400)}=-139~{\rm MeV} ~,
 \label{eq:fK1phys}
 \en
where we just consider the central values.

Finally, the $D \to A$ form factor parameters in the CLF quark model are given in Table~\ref{tab:FFDtoA}.

\begin{table}[t]
\caption{Parameters in the form factors of $D\to a_1(1260), b_1(1235), K_{1A}, K_{1B}$ transitions in the parametrization of Eq.~(\ref{eq:FFpara}), as obtained by fitting to the covariant light-front model \cite{CCH}. The numbers in parentheses are the form factors at $q^2=0$ obtained using the ISGW2 model \cite{ISGW2}. As noticed in the footnote of this subsection, the form factors for $D\to ^3P_1$ and $D\to ^1P_1$ transitions are of the same (opposite) signs in the CLF (ISGW) model.}
 \label{tab:FFDtoA}
\begin{ruledtabular}
\begin{tabular}{| c c c c || c c c c |}
~~~$F$~~~~~
    & $F(0)$~~~~~
    & $a$~~~~~
    & $b$~~~~~~
& ~~~ $F$~~~~~
    & $F(0)$~~~~~
    & $a$~~~~~
    & $b$~~~~~~
 \\
    \hline
$V^{Da_1}_0$
    & $0.31~(-0.60)$
    & $0.85$
    & $0.49$ &
$V_0^{Db_1}$
    & $0.49~(0.64)$
    & $0.89$
    & $0.28$
    \\
$V^{DK_{1A}}_0$
    & $0.34~(-0.37)$
    & $1.44$
    & $0.15$ &
$V^{DK_{1B}}_0$
    & $0.44~(0.50)$
    & $0.80$
    & $0.27$
    \\
\end{tabular}
\end{ruledtabular}
\end{table}

\subsection{Tensor mesons \label{sec:T}}

The observed $J^P=2^+$ tensor mesons $f_2(1270)$, $f_2'(1525)$, $a_2(1320)$ and $K_2^*(1430)$ form an SU(3) $1\,^3P_2$ nonet.  The $q\bar q$ content for isodoublet and isovector tensor resonances are obvious.  Just as the $\eta$-$\eta'$ mixing in the pseudoscalar case, the isoscalar tensor states $f_2(1270)$ and $f'_2(1525)$ also have a mixing, and their wave functions are defined by
 \be
 f_2(1270) &=&
{1\over\sqrt{2}}(f_2^u+f_2^d)\cos\theta_{f_2} + f_2^s\sin\theta_{f_2} ~, \non \\
 f'_2(1525) &=&
{1\over\sqrt{2}}(f_2^u+f_2^d)\sin\theta_{f_2} - f_2^s\cos\theta_{f_2} ~,
 \en
with $f_2^q\equiv q\bar q$.  Since $\pi\pi$ is the dominant decay mode of $f_2(1270)$ whereas $f_2'(1525)$ decays predominantly into $K\ov K$ (see Ref.~\cite{PDG}), it is obvious that this mixing angle should be small.  More precisely, it is found that $\theta_{f_2}=7.8^\circ$ \cite{Li} and $(9\pm1)^\circ$ \cite{PDG}.  Therefore, $f_2(1270)$ is primarily an $(u\bar u+d\bar d)/\sqrt{2}$ state, while $f'_2(1525)$ is dominantly $s\bar s$.

The polarization tensor $\vp_{\mu\nu}$ of a $^3P_2$ tensor meson with $J^{PC}=2^{++}$ satisfies the relations
 \be
 \vp_{\mu\nu}=\vp_{\nu\mu} ~, \qquad \vp^{\mu}_{~\mu}=0 ~, \qquad
 p_\mu \vp^{\mu\nu}=p_\nu\vp^{\mu\nu}=0 ~,
 \en
where $p^\mu$ is the momentum of the tensor meson.  Therefore,
 \be
 \la 0|(V-A)_\mu|T(\vp,p)\ra = a\vp_{\mu\nu}p^\nu+b\vp^\nu_{~\nu} p_\mu=0 ~,
 \en
and hence the decay constant of the tensor meson vanishes identically; that is, the tensor meson cannot be produced from the $V-A$ current.

The general expression for the $D\to T$ transition has the form \cite{ISGW}
 \be \label{DTff}
 \la T(\vp,p_T)|(V-A)_\mu|D(p_D)\ra &=&
 ih(q^2)\epsilon_{\mu\nu\rho\sigma}\vp^{*\nu\alpha}p_{D\alpha}(p_D+p_T)^\rho
 (p_D-p_T)^\sigma+k(q^2)\vp^*_{\mu\nu}p_D^\nu  \non \\
 &+& b_+(q^2)\vp^*_{\alpha\beta}p_D^\alpha p_D^\beta(p_D+p_T)_\mu
 +b_-(q^2)\vp^*_{\alpha\beta}p_D^\alpha p_D^\beta(p_D-p_T)_\mu.
 \en
The form factors $h$, $k$, $b_+$ and $b_-$ have been calculated in the ISGW quark model \cite{ISGW} and its improved version, the ISGW2 model \cite{ISGW2}.  They are also computed in the CLF quark model \cite{CCH} and listed in Table~\ref{tab:FFDtoT}.

The decay amplitude of $D\to TP$ always has the generic expression
 \be
 A(D\to TP)
 =\vp^*_{\mu\nu}p_D^\mu p_D^\nu\,M(D\to TP) ~.
 \en
The decay rate is given by
 \be \label{eq:rateTP}
 \Gamma(D\to TP)=\,{p_c^5\over 12\pi m_T^2}
 \left({m_D\over m_T}\right)^2|M(D\to TP)|^2 ~,
 \en
where $p_c$ is the magnitude of the 3-momentum of either final-state meson in the rest frame of the charmed meson.

\begin{table}[t]
\caption{Parameters in the form factors of $D\to a_2(132),K_2^*(1430)$ transitions in the parametrization of Eq.~(\ref{eq:FFpara}), as obtained by fitting to the covariant light-front model \cite{CCH}. The form factor $k$ is dimensionless, while $k$, $b_+$ and $b_-$ are in units
of ${\rm GeV}^{-2}$. The numbers in parentheses are the form factors at $q^2=0$ obtained using the ISGW2 model \cite{ISGW2}. }
 \label{tab:FFDtoT}
\begin{ruledtabular}
\begin{tabular}{| l c  c c || l c  c c |}
~$F$~~~~~
    & $F(0)$~~~~~
    & $a$~~~~~
    & $b$~~~~~~
& ~ $F$~~~~~
    & $F(0)$~~~~~
    & $a$~~~~~
    & $b$~~~~~~
 \\
    \hline
$h^{Da_2}$
    & $0.188~(0.203)$%$0.31$
    &  1.21
    &  1.09
&$k^{Da_2}$
    & $0.340~(0.613)$
    & $-0.07$
    & $0.12$
    \\
$b_+^{Da_2}$
    & $-0.084~(-0.052)$
    &  0.97
    &  0.58 &
$b_-^{Da_2}$
    & $0.120~(0.064)$
    & $1.15$
    & $0.66$\\
$h^{Df_{2q}}$
    & $0.17~(0.20)$%$0.31$
    &  1.28
    &  0.90
&$k^{Df_{2q}}$
    & $0.27~(0.61)$
    & $-0.21$
    & $0.12$
    \\
$b_+^{Df_{2q}}$
    & $-0.08~(-0.05)$
    &  1.02
    &  0.51 &
$b_-^{Df_{2q}}$
    & $0.10~(0.06)$
    & $1.14$
    & $0.49$\\
$h^{DK_2^*}$
    & $0.192~(0.14)$
    &  1.17
    &  0.99
&$k^{DK_2^*}$
    & $0.368~(0.71)$
    & $-0.04$
    & $0.11$
    \\
$b_+^{DK_2^*}$
    & $-0.096~(-0.060)$
    &  1.05
    &  0.58 &
$b_-^{DK_2^*}$
    & $0.137~(0.069)$
    & $1.17$
    & $0.69$
    \\
$h^{D_sf_{2s}}$
    & $0.15~(0.21)$
    &  1.04
    &  0.79
&$k^{D_sf_{2s}}$
    & $0.59~(1.15)$
    & $0.22$
    & $0.09$
    \\
$b_+^{D_sf_{2s}}$
    & $-0.09~(-0.09)$
    &  0.95
    &  0.54 &
$b_-^{D_sf_{2s}}$
    & $0.13~(0.12)$
    & $1.05$
    & $0.60$\\
$h^{D_sK_2^*}$
    & $0.15~(0.35)$
    &  1.11
    &  0.99
&$k^{D_sK_2^*}$
    & $0.42~(1.16)$
    & $0.08$
    & $0.11$
    \\
$b_+^{D_sK_2^*}$
    & $-0.07~(-0.08)$
    &  0.96
    &  0.60 &
$b_-^{D_sK_2^*}$
    & $0.11~(0.13)$
    & $1.10$
    & $0.63$
    \\
\end{tabular}
\end{ruledtabular}
\end{table}

\section{Diagrammatic  approach \label{sec:flavorapp}}

It has been established sometime ago that a least model-dependent analysis of heavy meson decays can be carried out in the so-called topological diagram approach. In this diagrammatic scenario, all two-body nonleptonic weak decays of heavy mesons can be expressed in terms of six distinct quark diagrams \cite{Chau,CC86,CC87}: $T$, the color-allowed external $W$-emission tree diagram; $C$, the color-suppressed internal $W$-emission diagram; $E$, the $W$-exchange diagram; $A$, the $W$-annihilation diagram; $P$, the horizontal $W$-loop diagram; and $V$, the vertical $W$-loop diagram.  (The one-gluon exchange approximation of the $P$ graph is the so-called ``penguin diagram''.)  It should be stressed that these diagrams are classified according to the topologies of weak interactions with all strong interaction effects encoded, and hence they are {\it not} Feynman graphs.  All quark graphs used in this approach are topological and meant to have all the strong interactions included, {\it i.e.}, gluon lines are included implicitly in all possible ways.  Therefore, analyses of topological graphs can provide information on FSIs.  Various topological amplitudes in two-body hadronic $D$ decays have been extracted from the data in \cite{Rosner99,Chiang03,Wu04,Wu05,Gronau94,RosnerPP08,RosnerVP,CLEOPP08,RosnerPP09} after making some reasonable approximations, {\it e.g.}, flavor SU(3) symmetry.

The topological amplitudes for $D\to SP,AP,TP$ decays have been discussed in \cite{ChengSP,ChengAP,ChengTP}.  There are several new features.  First, one generally has two sets of distinct external $W$-emission and internal $W$-emission diagrams, depending on whether the emitted particle is an even-party meson or an odd-parity one.  Let us denote the primed amplitudes $T'$ and $C'$ for the case when the emitted meson is an even-parity one.  Second, because of the smallness of the decay constants of even-parity mesons except for the $^3P_1$ axial-vector state, it is expected that $|T'|\ll |T|$ and $|C'|\ll |C|$. This feature can be tested experimentally.  Third, since $K^*_0$ and the light scalars $\sigma,~\kappa,~f_0,~a_0$ fall into two different SU(3) flavor nonets, in principle one cannot apply SU(3) symmetry to relate the topological amplitudes in $D^+\to f_0\pi^+$ to, for example, those in $D^+\to \ov K^{*0}_0\pi^+$.

\subsection{$D\to SP$ \label{sec:flavorDSP}}

{\squeezetable
\begin{table}[t]
\caption{Topological amplitudes and  branching fractions for various $D\to SP$ decays. In Scheme I, light scalar mesons $\sigma,~\kappa,~a_0(980)$ and $f_0(980)$ are described by the $q\bar q$  states, while $K^*_0$ as excited $q\bar q$ states. In Scheme II, light scalars are tetraquark states, while $K^*_0$ are ground-state $q\bar q$.  The $f_0-\sigma$ mixing angle $\theta$ in the 2-quark model is defined in Eq. (\ref{eq:mixing}). The experimental branching fractions for $D^0\to K_0^{*-}\pi^+$ and $D_s^+\to \ov K_0^{*0}K^+$ are taken from Table \ref{tab:DataSP} after average.
For simplicity, we do not consider the $f_0-\sigma$ mixing in the 4-quark model.}
\label{tab:DSP}
\begin{ruledtabular}
\begin{tabular}{l l l l}
Decay & Amplitude (I) & Amplitude (II) & $\B_{\rm expt}$ \\
 \hline
 $D^0\to f_0\pi^0$ & ${1\over 2}V_{cd}^*V_{ud}(-C+C'-E-E')\sin\theta$ & ${1\over 2}V_{cd}^*V_{ud}(-C+C'-E-E')$ &
 $(1.0\pm {0.3})\times 10^{-4}$ \\
 & \qquad $+{1\over\sqrt{2}}V_{cs}^*V_{us} C'\cos\theta$ & \qquad $ +V_{cs}^*V_{us} C'$ & \\
  \quad~ $\to f_0\ov K^0$ & $V_{cs}^*V_{ud}[{1\over\sqrt{2}}(C+E)\sin\theta+E'\cos\theta]$ & ${1\over\sqrt{2}}V_{cs}^*V_{ud}(C+2E'+E)$ & $({ 8.0}^{+2.5}_{-2.2})\times 10^{-3}$ \\
  \quad~ $\to a_0^+K^-$ & $ V_{cs}^*V_{ud}(T'+E)$ & $ V_{cs}^*V_{ud}(T'+E)$ & $$ \\
  \quad~ $\to a_0^0\ov K^0$ & $V_{cs}^*V_{ud}(C-E)/\sqrt{2}$  & $V_{cs}^*V_{ud}(C-E)/\sqrt{2}$ & $(1.6\pm0.5)\%$ \\
  \quad~ $\to a_0^-K^+$ & $ V_{cd}^*V_{us}(T+E')$ & $ V_{cd}^*V_{us}(T+E')$ & \\
  \quad~ $\to a_0^+\pi^-$ & $V_{cd}^*V_{ud}(T'+E)$ & $V_{cd}^*V_{ud}(T'+E)$ & $$ \\
  \quad~ $\to a_0^-\pi^+$ & $V_{cd}^*V_{ud}(T+E')$ & $V_{cd}^*V_{ud}(T+E')$ & $$ \\
  \quad~ $\to \sigma\pi^0$ & ${1\over 2}V_{cd}^*V_{ud}(-C+C'-E-E')\cos\theta$ & ${1\over \sqrt{2}}V_{cd}^*V_{ud}(-C+C'-E-E')$ & $({ 1.8\pm0.3})\times 10^{-4}$ \\
 & \qquad $-{1\over\sqrt{2}}V_{cs}^*V_{us} C'\sin\theta$ & \qquad $$ & \\
\hline
 $D^+\to f_0\pi^+$ & ${1\over\sqrt{2}}V_{cd}^*V_{ud}(T+C'+A+A')\sin\theta$ & ${1\over\sqrt{2}}V_{cd}^*V_{ud}(T+C'+A+A')$ &
 $({4.5}\pm 1.4)\times 10^{-4}$ \\
 & \qquad $+V_{cs}^*V_{us} C'\cos\theta$ & \qquad $+\sqrt{2}V_{cs}^*V_{us}C'$ & \\
 \qquad $\to f_0K^+$ &$V_{cd}^*V_{us}[{1\over\sqrt{2}}(T+A')\sin\theta+A\cos\theta]$ & ${1\over\sqrt{2}}V_{cd}^*V_{us}(T+2A+A')$ & $(1.6\pm1.0)\times 10^{-4}$ \\
 \qquad $\to a_0^+\ov K^0$ & $V_{cs}^*V_{ud}(T'+C)$ & $V_{cs}^*V_{ud}(T'+C)$ &  \\
 \qquad $\to a_0^0\pi^+$ & $V_{cd}^*V_{ud}(-T-C'-A+A')/\sqrt{2}$ & $V_{cd}^*V_{ud}(-T-C'-A+A')/\sqrt{2}$ & $$ \\
 \qquad $\to \sigma\pi^+$ & ${1\over\sqrt{2}}V_{cd}^*V_{ud}(T+C'+A+A')\cos\theta$  &  $V_{cd}^*V_{ud}(T+C'+A+A')$ & $(2.1\pm0.2)\times 10^{-3}$ \\
 & \qquad $-V_{cs}^*V_{us} C'\sin\theta$ & \qquad $$ & \\
%CC The following mode is a duplicate.
 \qquad $\to \bar\kappa^0K^+$ & $ V_{cs}^*V_{us}T + V_{cd}^*V_{ud}A$ & $ V_{cs}^*V_{us}T + V_{cd}^*V_{ud}A$ & $(1.0^{+0.5}_{-0.3})\times 10^{-3}$ \\
  \hline
 $D_s^+\to f_0\pi^+$ & $V_{cs}^*V_{ud}(T\cos\theta+(A+A')\sin\theta/\sqrt{2})$ & $ V_{cs}^*V_{ud}(2T+A+A')/\sqrt{2}$ & $(1.8\pm{ 0.5})\%$ \\
  \quad~ $\to f_0K^+$ & $V_{cs}^*V_{us}[(T+C'+A)\cos\theta+{1\over \sqrt{2}}A'\sin\theta]$ & ${1\over\sqrt{2}}V_{cs}^*V_{us}(2T+2C'+2A+A')$ &  $$ \\
  & \quad~$+{1\over\sqrt{2}}V_{cd}^*V_{ud} C'\sin\theta$ & $+{1\over\sqrt{2}}V_{cd}^*V_{ud}C'$ \\
\hline\hline
  $D^0\to K_0^{*-}\pi^+$ & $V_{cs}^*V_{ud}(T+E')$ & $V_{cs}^*V_{ud}(T+E')$ &   $(8.2\pm1.4)\times 10^{-3}$ \\
  \quad~ $\to \ov K_0^{*0}\pi^0$ & $V_{cs}^*V_{ud}(C'-E')/\sqrt{2}$ & $V_{cs}^*V_{ud}(C'-E')/\sqrt{2}$  &   $(9.2^{+8.1}_{ -2.6})\times 10^{-3}$  \\
%\hline
%
  $D^+\to \ov K_0^{*0}\pi^+$ & $V_{cs}^*V_{ud}(T+C')$  & $V_{cs}^*V_{ud}(T+C')$ &
$(1.8 \pm 0.3)\%$\footnotemark \\
%
%  \qquad $\to \ov K_0^{*0}K^+$ & $ V_{cs}^*V_{us}T + V_{cd}^*V_{ud}A$ & $ V_{cs}^*V_{us}T + V_{cd}^*V_{ud}A$ & $(3.0\pm0.6)\times 10^{-3}$ \\
%\hline
%
 $D_s^+\to \ov K_0^{*0}K^+$ & $V_{cs}^*V_{ud}(C'+A)$ & $V_{cs}^*V_{ud}(C'+A)$ & $(3.6\pm0.7)\times 10^{-3}$ \\
  \quad~ $\to K_0^{*0}\pi^+$ & $V_{cd}^*V_{ud}\,T+V_{cs}V_{us}^*\,A$ &  $V_{cd}^*V_{ud}\,T+V_{cs}V_{us}^*\,A$ & $(8.1\pm6.5)\times 10^{-4}$  \\
\end{tabular}
\end{ruledtabular}
\footnotetext[1]{Data from E791 \cite{E791}; see also Eq. (\ref{eq:E791}).}
\end{table}}

The topological amplitudes for $D\to SP$ decays are listed in Table~\ref{tab:DSP} for two different schemes. In Scheme I, light scalar mesons $\sigma, \kappa, a_0(980)$ and $f_0(980)$ are described by the ground-state $q\bar q$  states, while $K^*_0$ as excited $q\bar q$ states. In Scheme II, light scalars are tetraquark states, while $K^*_0$ are ground-state $q\bar q$. The expressions of topological amplitudes are the same in both Schemes I and II except for the channels involving $f_0$ or $\sigma$.

Since the decay constant of $f_0$ and $\sigma$ vanishes, one can set $T' = C'=0$. From Table~\ref{tab:DSP} we have
\be
\label{eq:schemedep}
{\B(D^+\to f_0\pi^+)\over \B(D^+\to \sigma\pi^+)}={\B(D^0\to f_0\pi^0)\over \B(D^0\to \sigma\pi^0)} \simeq \left\{
    \begin{array}{cl}
     \tan^2\theta
      & \quad \mbox{2-quark,}  \\
     {1\over 2}
      & \quad \mbox{4-quark.} \end{array}\right.
\en
It appears that the data of $D^+\to f_0\pi^+,\sigma\pi^+$ favor the 2-quark picture of the light scalars, while the measurements of $D^0\to f_0\pi^0,\sigma\pi^0$ prefer the 4-quark scenario.  Moreover, the $D^0 \to a_0^+ K^-$ and $a_0^+ \pi^-$ modes will be dominated by the $W$-exchange diagram, $E$.  The $D^+ \to a_0^+ \ov K^0$ mode is dominated by the $C$ amplitude.  The $D^0 \to \ov K_0^{*0} \pi^0$, $D^+ \to K_0^{*0} \pi^+$, and $D_s^+ \to K_0^{*0} K^+$ are dominated by the $E'$, $T$, and $A$ amplitudes, respectively.

\begin{table}[t]
\caption{Extracted flavor amplitude parameters from fits to the $D \to S P$ decays, where $S$ only refers to lighter scalar mesons here.  The amplitude magnitudes are in units of $10^{-6}$ GeV.  In these fits, we set $C' = 0$, $E' = E$, and $A' = -A$, with reasons explained in the text.  The strong phases $\delta_{E,A}$ are associated with the $E$ and $A$ amplitudes, respectively.  For Fits (A) and (B), we take the mixing angle $\theta = 25^\circ$.}
\label{tab:DSPfit}
\begin{ruledtabular}
\begin{tabular}{ccc|cc}
Scheme & \multicolumn{2}{c}{I} & \multicolumn{2}{c}{II} \\
Fit & (A) & (B) & (A) & (B) \\
\hline
$|T|$ & $2.14^{+0.16}_{-0.15}$ & --- & $1.55 \pm 0.07$ & --- \\
$|A|$ & $3.16^{+0.10}_{-0.11}$ & --- & $2.15^{+0.30}_{-0.45}$ & --- \\
$\delta_{A}$ & $(31 \pm 2)^\circ$ & --- & $(35^{+9}_{-11})^\circ$ & --- \\
\hline
$|C|$ & --- & $1.44^{+0.15}_{-0.14}$ & --- & $1.90^{+0.36}_{-0.22}$ \\
$|E|$ & --- & $1.20^{+0.05}_{-0.06}$ & --- & $1.18 ^{+0.05}_{-0.06}$ \\
$\delta_{E}$ & --- & $(168 ^{+35}_{-10})^\circ$ & --- & $(152^{+3}_{-2})^\circ$ \\
\hline
$\chi^2_{\rm min}$ / d.o.f & $2.61 / 2$ & $5.07 / 1$ & $8.29 / 2$ & $0.74 / 1$
\end{tabular}
\end{ruledtabular}
\end{table}

Table~\ref{tab:DSP} is divided into two parts separated by double lines.  The upper part involves only light scalar mesons ($f_0$, $a_0$, $\sigma$, and $\kappa$), whereas the lower part involves the $K_0^*$ mesons in the heavier nonet representation.  This division is made because the amplitudes of the same topology in these two groups have no {\it a priori} relations.  We first note that none of the currently measured modes involve the $T'$ amplitude.  Secondly, one can simplify the upper part of the table by setting $C' = 0$, for the decay constants of scalar mesons are expected to be either identically zero or relatively small.  Moreover, the modes in the lower part of Table~\ref{tab:DSP} have the same amplitude decomposition in the two schemes and involve the $T$, $C'$, $E'$, and $A$ amplitudes.  One cannot set $C' = 0$ here because the decay constant of $K_0^*$ is non-negligible, as commented after Eq.~(\ref{eq:EOM}).  In this case, there are more theory parameters than observables, barring a fit.

In the following, we will perform two sets of fit [(A) and (B)] to the modes involving only the lighter scalar mesons in the flavor diagram formalism.  Fits (A) includes the five measured $D^+$ and $D_s^+$ decays.  As noted in Eq.~(\ref{eq:schemedep}), the $f_0 \pi^+$ and $\sigma \pi^+$ modes are related because of the same flavor amplitude combination.  Here we have to assume a relation between $A$ and $A'$ in order to reduce the number of parameters.  Without further theoretical guidance, we have tried the cases $A' = A$ and $A' = -A$ for simplicity and found that the latter renders an equally good or better fit than the former.  The strong phase $\delta_A$ in Table~\ref{tab:DSPfit} is measured with respect to $T$, which is assumed real.  We also note here that all the strong phases given in the table are subject to a two-fold ambiguity ($\delta \to -\delta$).  From the $\chi^2_{\rm min}$ values, one sees that Scheme I fits better than Scheme II in these modes.  This is understandable because the 2-quark picture explains better the observed rates of the $f_0 \pi^+$ and $\sigma \pi^+$ decays.  In either scheme, $|A|$ is about $1.5$ times larger than $|T|$, showing the importance of the $W$-annihilation contribution.  Besides, the extracted relative strong phase is robust.

Fit (B) includes the four measured $D^0$ decays.  Here it does not matter what relation we assume between $E$ and $E'$, as far as the $\chi^2_{\rm min}$ value is concerned.  The only effect is on the size and phase of the $E$ amplitude.  This is because the $f_0 \pi^0$ and $\sigma \pi^0$ modes are related by the same flavor amplitude combination, as also noted in Eq.~(\ref{eq:schemedep}).  We assume $E' = E$ for an explicit fit.  In Table~\ref{tab:DSPfit}, the strong phase $\delta_E$ is measured with respect to $C$, which is assumed real.  The $\chi^2_{\rm min}$ values show that Scheme II explains this set of data better, as noted below Eq.~(\ref{eq:schemedep}).  It is worth noting that the magnitudes and relative phase extracted in either scheme are roughly the same.  Also, the amplitudes $C$ and $E$ are almost opposite in phase, as required primarily by the $D^0 \to a_0^0 \ov K^0$ decay.  Finally, based on the $\chi^2_{\rm min}$ values of these fits, the current data still cannot differentiate the two schemes yet.

\subsection{$D\to AP$ \label{sec:flavorDAP}}

\begin{table}[t]
\caption{Topological amplitudes and branching fractions of $D\to AP$ decays.  The notation is explained in the main text.  Theory predictions are made within the factorization approach, with the mixing angle $\theta_{K_1} = 50.8^\circ$.} \label{tab:DtoAPtheory}
\begin{ruledtabular}
\begin{tabular}{ l c c c }
Decay & Amplitude & Theory & Experiment \\
    \hline
$D^+\to \bar K_1^0(1270)\pi^+$
    & $V_{cs}^* V_{ud}
    \left[ (T_A+C'_A)\sin\theta_{K_1} + (T_B+C'_B)\cos\theta_{K_1} \right]$
    & $4.7\times 10^{-3}$
    & $<7\times 10^{-3}$
    \\
$D^+\to \bar K_1^0(1400)\pi^+$
    & $V_{cs}^* V_{ud}
    \left[ (T_A+C'_A)\cos\theta_{K_1} - (T_B+C'_B)\sin\theta_{K_1} \right]$
    & $2.2\%$
    & $(3.8\pm1.3)\%$
    \\
$D^+\to \bar K^0 a_1^+(1260)$
    & $V_{cs}^* V_{ud} (T'_A+C_A)$
    & $8.2\%$
    & $(7.0\pm1.2)\%$
    \\
$D^+\to \bar K^0 b_1^+(1235)$
    & $V_{cs}^* V_{ud} (T'_B+C_B)$
    & $2.2\times 10^{-3}$
    & $$
    \\
\hline
$D^0\to K_1^-(1270)\pi^+$
    & $V_{cs}^* V_{ud}
    \left[ (T_A+E'_A)\sin\theta_{K_1} + (T_B+E'_B)\cos\theta_{K_1} \right]$
    & $5.2\times 10^{-3}$
    & $(1.14\pm0.32)\%$
    \\
$D^0\to K_1^-(1400)\pi^+$
    & $V_{cs}^* V_{ud}
    \left[ (T_A+E'_A)\cos\theta_{K_1} - (T_B+E'_B)\sin\theta_{K_1} \right]$
    & $1.4\times 10^{-4}$
    & $<1.2\%$
    \\
$D^0\to \bar K_1^0(1270)\pi^0$
    & $V_{cs}^* V_{ud} {1\over\sqrt{2}}
    \left[ (C'_A-E'_A)\sin\theta_{K_1} + (C'_B-E'_B)\cos\theta_{K_1} \right]$
    & $6.6\times 10^{-3}$
    & $$
    \\
$D^0\to \bar K_1^0(1400)\pi^0$
    & $V_{cs}^* V_{ud} {1\over\sqrt{2}}
    \left[ (C'_A-E'_A)\cos\theta_{K_1} - (C'_B-E'_B)\sin\theta_{K_1} \right]$
    & $3.2\times 10^{-3}$
    & $<3.7\%$
    \\
$D^0\to K_1^+(1270)K^-$
    & $V_{cs}^* V_{us}
    \left[ (T'_A+E_A) \sin\theta_{K_1} + (T'_B+E_B)\cos\theta_{K_1} \right]$
    & $4.6\times 10^{-4}$
    & $$
    \\
$D^0\to K_1^-(1270)K^+$
    & $V_{cs}^* V_{us}
    \left[ (T_A+E'_A) \sin\theta_{K_1} + (T_B+E'_B)\cos\theta_{K_1} \right]$
    & $8.2\times 10^{-5}$
    & $$
    \\
$D^0\to K_1^\pm(1270)K^\mp$
    &
    & $5.4\times 10^{-4}$
    & $(8.1\pm1.8)\times 10^{-4}$
    \\
$D^0\to K^- a_1^+(1260)$
    & $V_{cs}^* V_{ud} (T'_A+E_A)$
    & $2.7\%$
    & $(7.9\pm1.1)\%$
    \\
$D^0\to \bar K^0 a_1^0(1260)$
    & $V_{cs}^* V_{ud} {1\over\sqrt{2}}(C_A-E_A)$
    & $1.2\times 10^{-4}$
    & $<1.9\%$
    \\
$D^0\to \pi^- a_1^+(1260)$
    & $V_{cd}^* V_{ud} (T'_A+E_A)$
    & $5.1\times 10^{-3}$
    & $(8.98\pm0.62)\times 10^{-3}$
    \\
$D^0\to K^- b_1^+(1235)$
    & $V_{cs}^* V_{ud} (T'_B+E_B)$
    & $1.7\times 10^{-5}$
    & $$
    \\
$D^0\to \bar K^0 b_1^0(1235)$
    & $V_{cs}^* V_{ud} {1\over\sqrt{2}}(C_B-E_B)$
    & $3.0\times 10^{-4}$
    &
    \\
\end{tabular}
\end{ruledtabular}
\end{table}

The topological amplitudes for $D \to AP$ decays are given in Table~\ref{tab:DtoAPtheory}.  Instead of using subscripts of $A$ and $P$ to complicate the notation, we use the primed (unprimed) amplitudes to indicate that the spectator quark in the $D$ meson ends up in the pseudoscalar (axial-vector) meson in the final state.  The subscripts $A$ and $B$ refer to the amplitudes associated with the $^3P_1$ and $^1P_1$ axial-vector mesons, respectively.  However, as we will see later in Section~\ref{sec:facDAP}, the factorization approach predicts that such a distinction is only necessary for the $T$ amplitudes.  The assumption of $C_A = C_B$ and $E_A = E_B$ can be checked by comparing the rates of $D^0 \to {\bar K}^0 a_1^0(1260)$ and $D^0 \to {\bar K}^0 b_1^0(1235)$, which are seen to be roughly the same up to a tiny phase space difference.  With the flavor symmetry assumption, the magnitudes of the invariant amplitudes of $D^0 \to \pi^- a_1^+(1260)$ and $D^0 \to K^- a_1^+(1260)$ should differ by a factor of $\lambda \simeq 0.2253$, which is to be compared with $0.171 \pm 0.013$ given by the current data.  A distinctive feature between the Cabibbo-allowed $D^0$ and $D^+$ decays is that the $W$-exchange diagrams ($E$) only involve in the former.

Current data for $D \to AP$ decays (only six branching fractions) are still insufficient for a sensible fit.  The theory predictions in Table~\ref{tab:DtoAPtheory} are based on the factorization calculations given in Section~\ref{sec:facDAP}.

\subsection{$D\to TP$ \label{sec:flavorDTP}}

\begin{table}[t]
\caption{Topological amplitudes and branching fractions of $D\to TP$ decays.  The notation is explained in the main text.  Theory predictions are made within the factorization approach, with the mixing angle $\theta_{f_2} = 7.8^\circ$.} \label{tab:DtoTPtheory}
\begin{ruledtabular}
\begin{tabular}{ l c c c }
Decay & Amplitude & Theory & Experiment \\
\hline
 $D^+\to f_2\pi^+$
& $\frac{1}{\sqrt{2}}V_{cd}^*V_{ud}\cos\theta_{f_2} (T + C' + A + A')$
& $0.9\times 10^{-6}$ & $(8.8\pm 1.6)\times 10^{-4}$ \\
& $+ V_{cs}^*V_{us}\sin\theta_{f_2} C'$ & \\
 $D^+\to a_2^+\ov K^0$
& $V_{cs}^*V_{ud} (T' + C)$
& $4.5\times 10^{-7}$ & $<3.0\times 10^{-3}$ \\
 $D^+\to \ov K^{*0}_2\pi^+$
& $V_{cs}^*V_{ud} (T + C')$
& $1.9\times 10^{-5}$ & $(6.3\pm 1.2)\times 10^{-4}$ \\
 $D^+\to K^{*0}_2\pi^+$
& $V_{cd}^*V_{us} (C' + A)$
& 0 & $(1.5\pm 1.0)\times 10^{-4}$ \\
 \hline
 $D^0\to f_2\pi^0$
& $\frac12 V_{cd}^*V_{ud}\cos\theta_{f_2} (C' - C - E' - E)$
& $5.1\times 10^{-8}$ & $(3.4\pm 0.4)\times 10^{-4}$ \\
& $+ \frac{1}{\sqrt{2}} V_{cs}^*V_{us}\sin\theta_{f_2} C'$\\
 $D^0\to f_2\ov K^0$
& $V_{cs}^*V_{ud} \left[
\frac{1}{\sqrt{2}}\cos\theta_{f_2} (C + E) + \sin\theta_{f_2} E' \right]$
& $1.5\times 10^{-7}$ & $(5.0^{+3.5}_{-2.1})\times 10^{-4}$ \\
 $D^0\to a_2^+K^-$
& $V_{cs}^*V_{ud} (T' + E)$
& 0 & $<2\times 10^{ -3}$ \\
 $D^0\to K^{*-}_2\pi^+$
& $V_{cs}^*V_{ud} (T + E')$
& $7.5\times 10^{-6}$ & $(2.1^{+1.2}_{ -0.7})\times 10^{-3}$ \\
\hline
 $D_s^+\to f_2\pi^+$
& $V_{cs}^*V_{ud} \left[
\frac{1}{\sqrt{2}}\cos\theta_{f_2} (A + A') + \sin\theta_{f_2} T \right]$
& $7.0\times 10^{-6}$ & $(1.9\pm 0.4)\times 10^{-3}$ \\
%
% $D_s^+\to f_2 K^+$
%& $\frac{1}{\sqrt{2}}\cos\theta_{f_2} (V_{cd}^*V_{ud} C' + V_{cs}^*V_{us} A')$
%& $3.1\times 10^{-8}$ & $(3.5\pm 2.3)\times 10^{-4}$ \\
%& $+ V_{cs}^*V_{us} \sin\theta_{f_2} (T + C' + A)$ \\
%
 $D_s^+\to K_2^{*0}\pi^+$
& $V_{cd}^*V_{ud} T + V_{cs}^*V_{us} A$
& $2.4\times 10^{-6}$ & $(1.5\pm1.2)\times 10^{-3}$  \\
\end{tabular}
\end{ruledtabular}
\end{table}

The topological amplitudes for $D \to TP$ decays are given in Table~\ref{tab:DtoTPtheory}.  Here we also use the unprimed (primed) symbols to indicate that the spectator quark of the $D$ meson ends up in the even-parity (tensor) and odd-parity (pseudoscalar) mesons in the final state, respectively.  There should be no confusion even though the amplitude symbols used here are identical to those in the $D \to SP$ case.

As described before, the decay constants of the tensor mesons vanish identically.  Therefore, one can set $T' = C' = 0$ in Table~\ref{tab:DtoTPtheory}.  The decay $D^+ \to \ov{K_2}^{*0} K^+$ is kinematically forbidden as the $K_2^*$ mass is above the kinematic threshold, though it is physically allowed through the width of $K_2^*$.  We therefore will not include it in our fit.  Note that the measured $D^+$ and $D_s^+$ decays only involve the $T$, $A$, and $A'$ amplitudes.  The results of fits to these decay modes are given in Table~\ref{tab:DTPfit}.  Since there are only three measured modes in the $D^0$ decays whereas at least four additional parameters have to be introduced even if we set $E' = E$, it is impossible to determine the magnitudes and strong phases of $C$ and $E$ by considering a global fit to the $TP$ decays.  For that, a determination of ${\cal B}(D^0 \to a_2^+ K^-)$ is crucial.

\begin{table}[t]
\caption{Extracted flavor amplitude parameters from fits to the $D \to TP$ decays.  The amplitude magnitudes are in units of $10^{-6} {\rm GeV}^{-1}$, the same as $M$ in Eq.~(\ref{eq:rateTP}).  In these fits, we set $C' = 0$ and $\theta_{f_2} = 7.8^\circ$.  The strong phases $\delta_{A,A'}$ are associated with the $A$ and $A'$ amplitudes, respectively, relative to $T$.  The contents of different fits are described in the main text.}
\label{tab:DTPfit}
\begin{ruledtabular}
\begin{tabular}{lllllll}
Parameter & $|T|$ & $|A|$ & $\delta_A$
& $|A'|$ & $\delta_{A'}$ & $\chi^2_{\rm min}$ / d.o.f. \\
\hline
Fit (A) & $1.85^{+0.14}_{-0.16}$ & $12.63^{+0.15}_{-0.16}$ & $(0 \pm 3)^\circ$
& $11.24^{+0.16}_{-0.15}$ & $(180 \pm 3)^\circ$ & $11.4 / 0$ \\
Fit (B) & $8.13^{+0.39}_{-0.42}$ & $1.59 \pm 0.08$ & $(173^{+26}_{-12})^\circ$
& --- & --- & $0 / 0$ \\
\end{tabular}
\end{ruledtabular}
\end{table}

Fit (A) in Table~\ref{tab:DTPfit} includes all the available $D^+$ and $D_s^+$ decay modes except for $D^+ \to \ov{K_2}^{*0} K^+$; there are thus 5 observables for 5 parameters.  According to the results of Fit (A), the current data favor relatively large $W$-annihilation diagrams.  This is because $|A|$ and $|A'|$ are constrained respectively by the doubly Cabibbo-suppressed $D^+ \to K_2^{*0} \pi^+$ mode and the singly Cabibbo-suppressed $D_s^+ \to f_2 K^+$ mode to be large.  Moreover, $A$ and $A'$ are about the same size but opposite in phase so that the branching fraction of $D_s^+ \to f_2 \pi^+$ falls in the ball park.  On the other hand, $|T|$ is largely constrained by the Cabibbo-favored $D^+ \to \ov{K_2}^{*0} \pi^+$ mode to be small.  The largest contribution in the $\chi^2_{\rm min}$ value comes from the $D^+ \to f_2 \pi^+$ mode ($\sim 10.1$).  This is a manifestation of the disparity between the Cabibbo-favored $D_s^+ \to f_2 \pi^+$ decay and the singly Cabibbo-suppressed $D^+ \to f_2 \pi^+$ decay that are seen to have similar branching ratios, if $|T|$ is constrained not to play a role here.

In view of the possibly problematic $D^+ \to K_2^{*0} \pi^+$ and $\ov{K_2}^{*0} \pi^+$ modes, we exclude them in Fit (B) and set $A' = A$ for simplicity; there are then 3 observables for 3 parameters.  We note that it is not illuminating to consider $A' = -A$ here because two of these modes involve the combination $A+A'$ as the major contribution.  In Fit (B), $|T|$ becomes larger and $A$ much suppressed.  Also, the relative strong phase between $A$ and $T$ is almost opposite to that in Fit (A).

The third column in Table~\ref{tab:DtoTPtheory} lists theory predictions based on factorization assumption to be discussed in Section~\ref{sec:facDTP}.  A comparison between the predictions and the measured values shows an apparent deficit in theory account of the decay amplitudes.  First, the magnitude of the tree contribution in the factorizationa approach is even smaller than the value of $|T|$ in Table~\ref{tab:DTPfit}.  Secondly, it is necessary to invoke the annihilation type of amplitudes to explain the observed data.

\section{Factorization Approach \label{sec:facapp}}

The diagrammatic approach has been applied quite successfully to hadronic decays of charmed mesons into $PP$ and $VP$ final states \cite{Rosner99,Chiang03,Wu04,Wu05,RosnerPP08,RosnerVP,CLEOPP08,RosnerPP09}.
When generalized to the decay modes involving an even-parity light meson in the final state, it appears that the current data are still insufficient for us to fully extract the information of all amplitudes.  Moreover, as shown in Tables~\ref{tab:DSPfit} and \ref{tab:DTPfit}, the extracted parameters do not present a coherent picture yet.  Therefore, we take the naive factorization formalism as a complementary approach to estimate the rates of these decay modes.  In this framework, the $W$-exchange and -annihilation type of cotributions will be neglected.  We discuss the three categories of decays in the following subsections separately.

\subsection{$D\to SP$ \label{sec:facDSP}}

The factorizable amplitudes for the $D\to SP$ decays involve
 \begin{eqnarray} \label{eq:XDSP}
  X^{(D S, P)}
 &=& \langle P(q)| (V-A)_\mu|0\rangle \langle S(p)| (V-A)^\mu|D(p_D)\rangle, \non \\
  X^{(D P, S)}
 &=& \langle S(q)| (V-A)_\mu|0\rangle \langle P(p)| (V-A)^\mu|D(p_D)\rangle,
 \end{eqnarray}
with the expressions
 \be \label{eq:XSP}
 X^{(DS, P)}
 =  -f_P(m_D^2-m_S^2) F_0^{DS}(q^2)\,,  \qquad
  X^{(D P, S)}=
 f_S (m_D^2-m_P^2) F_1^{DP}(q^2)\,,
 \end{eqnarray}
where use of Eqs.~(\ref{eq:Sdecayc}) and (\ref{DSm.e.}) has been made. The decay amplitudes of $D\to K_0^*P$ thus read
 \be \label{eq:SPamp}
 A(D^+\to \ov K^{*0}_0\pi^+) &=&
 {G_F\over\sqrt{2}}V_{cs}^*V_{ud}\Big
 [-a_1f_\pi(m_D^2-m_{K^*_0}^2)F_0^{DK^*_0}(m_\pi^2)  \non \\
 && \quad + a_2 f_{K^*_0}(m_D^2-m_\pi^2)F_0^{D\pi}(m_{K^*_0}^2)\Big] ~.\non \\
 A(D^0\to K^{*-}_0\pi^+) &=& -{G_F\over\sqrt{2}}V_{cs}^*V_{ud}\,a_1f_\pi(m_D^2-m_{K^*_0}^2)
F_0^{DK_0^*}(m_\pi^2) ~,  \non \\
 A(D^0\to \ov K_0^{*0}\pi^0) &=& {G_F\over
 2}V_{cs}^*V_{ud} \,a_2f_{K^*_0}(m_D^2-m_\pi^2)F_0^{D\pi}(m^2_{K^*_0}) ~,
 \non \\
  A(D^+_s\to K_0^{*0}\pi^+) &=& -{G_F\over \sqrt{2}}\,a_1V_{cd}^*V_{ud}
  f_\pi(m_{D_s}^2-m_{K^*_0}^2)F_0^{D_sK^*_0}(m^2_\pi) ~,
 \en
and likewise for the other $D\to SP$ decays.

\begin{table}[t]
\caption{The predicted branching fractions for various $D\to SP$ decays with the scalar mesons treated as $q\bar q$ ground states.  For simplicity, we have dropped the mass identification for $f_0(980)$ and $K^*_0(1430)$. The $f_0-\sigma$ mixing angle $\theta$ is taken to be $25^\circ$. Theory predictions are made within the factorization approach in which the weak annihilation topologies ($E$ and $A$) are neglected.}
\label{tab:DtoSPtheory}
\begin{ruledtabular}
\begin{tabular}{l l l l}
Decay & Amplitude & $\B_{\rm theory}$ & $\B_{\rm expt}$ \\
 \hline
 $D^0\to f_0\pi^0$ & ${1\over 2}V_{cd}^*V_{ud}(-C+C'-E-E')\sin\theta$ & $7.8\times 10^{-6}$ &
 $(1.0\pm {0.3})\times 10^{-4}$ \\
 & \qquad $+{1\over\sqrt{2}}V_{cs}^*V_{us} C'\cos\theta$ & & \\
  \quad~ $\to f_0\ov K^0$ & $V_{cs}^*V_{ud}[{1\over\sqrt{2}}(C+E)\sin\theta+E'\cos\theta]$ & $3.5\times 10^{-4}$ & $({ 8.0}^{+2.5}_{-2.2})\times 10^{-3}$ \\
 $D^+\to f_0\pi^+$ & ${1\over\sqrt{2}}V_{cd}^*V_{ud}(T+C'+A+A')\sin\theta$ & $1.4\times 10^{-4}$ &
 $({4.5}\pm 1.4)\times 10^{-4}$ \\
 & \qquad $+V_{cs}^*V_{us} C'\cos\theta$ & \qquad $$ & \\
 \qquad $\to f_0K^+$ &$V_{cd}^*V_{us}[{1\over\sqrt{2}}(T+A')\sin\theta+A\cos\theta]$ & $1.1\times 10^{-5}$ & $(1.6\pm1.0)\times 10^{-4}$ \\
 $D_s^+\to f_0\pi^+$ & $V_{cs}^*V_{ud}(T\cos\theta+(A+A')\sin\theta/\sqrt{2})$ & $1.3\%$ & $(1.8\pm{ 0.5})\%$ \\
\hline
  $D^0\to K_0^{*-}\pi^+$ & $V_{cs}^*V_{ud}(T+E')$ & $2.0\times 10^{-3}$ &   $(8.2\pm1.4)\times 10^{-3}$ \\
  \quad~ $\to \ov K_0^{*0}\pi^0$ & $V_{cs}^*V_{ud}(C'-E')/\sqrt{2}$ & $1.2\times 10^{-3}$  &   $(9.2^{+8.1}_{ -2.6})\times 10^{-3}$  \\
  $D^+\to \ov K_0^{*0}\pi^+$ & $V_{cs}^*V_{ud}(T+C')$  & $2.3\%$ & $(1.8\pm0.3)\%$ \footnotemark[1]\\
 $D_s^+\to \ov K_0^{*0}K^+$ & $V_{cs}^*V_{ud}(C'+A)$ & $5.5\times 10^{-4}$ & $(3.6\pm0.7)\times 10^{-3}$ \\
  \quad~ $\to K_0^{*0}\pi^+$ & $V_{cd}^*V_{ud}\,T+V_{cs}V_{us}^*\,A$ &  $2.6\times 10^{-4}$ & $(8.1\pm6.5)\times 10^{-4}$  \\
\end{tabular}
\footnotetext[1]{Data from E791 \cite{E791}; see also Eq. (\ref{eq:E791}).}
\end{ruledtabular}
\end{table}

Using the decay constants and form factors given in Section~\ref{sec:S}, the predicted rates of $(D,D_s)\to (f_0,K_0^*)P$ are computed and listed in Table~\ref{tab:DtoSPtheory}, where we have used $a_1=1.22$, $a_2=-0.66$ and taken the form factors for $D$ to $\pi$ and $K$ transitions from the recent CLEO-c measurements of semileptonic $D$ meson decays to $\pi$ and $K$ mesons \cite{CLEO:FF}. In order to test the factorization approach, we should focus on the modes in which weak annihilations ($W$-exchange or $W$-annihilation) are absent or suppressed. The Cabibbo-allowed decays $D^+\to \ov K_0^{*0}\pi^+$ and $D_s^+\to f_0\pi^+$ satisfy this criterion: the weak annihilation amplitude is absent in the former and suppressed by the $f_0-\sigma$ mixing in the latter. We see from Table~\ref{tab:DtoSPtheory} that factorization works well for these two modes. For Cabibbo-allowed $D^+\to PP$ or $VP$ decays, it is known that the color-allowed $T$ and color-suppressed $C$ amplitudes interfere destructively due to the opposite sign of the parameters $a_1$ and $a_2$. However, it is the other way around for Cabibbo-allowed $D^+\to SP$ and $AP$ decays. From Eq.~(\ref{eq:SPamp}), it is obvious that the $a_1$ and $a_2$ terms in the decay amplitude of $D^+\to \ov K_0^{*0}\pi^+$ interfere constructively. If they interfered destructively, one would have $\B(D^+\to \ov K_0^{*0}\pi^+)=5.4\times 10^{-5}$ which is too small compared to experiment. Numerically, we obtain $|T|=9.2\times 10^{-7}$ GeV and $|C'|=1.0\times 10^{-6}$ GeV for this mode. Therefore, even though $C'$ is suppressed by the smallness of $f_{K_0^*(1430)}$ and $a_2$, it is enhanced sizably by the mass squared term $(m_D^2-m_\pi^2)$ and the form factor $F_0^{D\pi}$ at $q^2=m_{K_0^*}^2$.

From Table~\ref{tab:DtoSPtheory} we see that the predicted rates for the other $D\to \ov K_0^*P$ ($P=\pi,K$) decays are smaller than experiments by a factor of $2\sim 8$. Note that they always receive weak annihilation contributions ($E$ or $A$). Under the factorization hypothesis, the factorizable $W$-exchange and $W$-annihilation amplitudes are suppressed due to the smallness of the form factor at large $q^2=m_D^2$.  This corresponds to the so-called helicity suppression. However, sizable long-distance weak annihilation can be induced via FSIs.  For charm decays, it is expected that the long-distance weak annihilation is dominated by resonant FSIs.  That is, the FSI via $q\bar q$ resonances is usually the most important one due to the fact that an abundant spectrum of resonances is known to exist at energies close to the masses of the charmed mesons. The diagrammatic-approach analysis in the last section suggests that weak annihilation diagrams are comparable to or even larger than the color-allowed tree amplitude $T$. Therefore, it is conceivable that the inclusion of $E'$ and $A$ amplitudes can account for ${\cal B}(D\to \ov K_0^*P)$. For $D(D_s^+)\to f_0\pi$ and $f_0K$ decays, the calculated branching fractions are typically too small by about one order of magnitude.  To enhance the rates in this case, the weak annihilation contributions have to be larger than the color-allowed tree amplitude, as shown in Table~\ref{tab:DSPfit}.  Such an amplitude hierarchy poises a difficulty in theoretical understanding.  While the non-strange content of $f_0(980)$ is small in the two-quark model for light scalars, it is not so in the tetraquark picture. This suggests that one should treat the light scalar mesons as bound states of $qq\bar q\bar q$. Unfortunately, the naive quark model is not applicable to evaluating the form factors for the transition of $D$ to a 4-quark state.

\subsection{$D\to AP$ \label{sec:facDAP}}

The factorizable amplitudes for the $D\to AP$ decays involve
 \begin{eqnarray} \label{eq:XDAP}
  X^{(D A, P)}
 &=& \langle P(q)| (V-A)_\mu|0\rangle \langle A(p)| (V-A)^\mu|D(p_D)\rangle, \non \\
  X^{(D P, A)}
 &=& \langle A(q)| (V-A)_\mu|0\rangle \langle P(p)| (V-A)^\mu|D(p_D)\rangle.
 \end{eqnarray}
with the expressions
 \be
 X^{(DA, P)}
 =  2  f_P m_{A} V_0^{DA}(q^2) (\epsilon^{*}\cdot p_D) ~,  \qquad
  X^{(D P, A)}=
 - 2  f_A m_{A} F_1^{DP}(q^2) (\epsilon^{*}\cdot p_D) ~.
 \end{eqnarray}
It is then straightforward to write down the factorizable amplitudes of $D\to K_1(1270)\pi$ and $D\to K_1(1400)\pi$ decays (dropping the overall $\vp^*\cdot p_D$ terms for simplicity):
 \be \label{eq:K1pi}
 A(D^+\to \ov K^0_1(1270)\pi^+) &=&
 {G_F\over\sqrt{2}}V_{cs}^*V_{ud}\Big[2a_1m_{K_1(1270)}f_\pi(\sin\theta_{K_1}
 V_0^{DK_{1A}}(m_\pi^2)+\cos\theta_{K_1} V_0^{DK_{1B}}(m_\pi^2))  \non \\
 &-& 2a_2m_{K_1(1270)}f_{K_1(1270)}F_1^{D\pi}(m^2_{K_1(1270)})\Big] ~, \non \\
 A(D^+\to \ov K^0_1(1400)\pi^+) &=&
 {G_F\over\sqrt{2}}V_{cs}^*V_{ud}\Big[2a_1m_{K_1(1400)}f_\pi(\cos\theta_{K_1}
 V_0^{DK_{1A}}(m_\pi^2)-\sin\theta_{K_1} V_0^{DK_{1B}}(m_\pi^2))  \non \\
 &-& 2a_2m_{K_1(1400)}f_{K_1(1400)}F_1^{D\pi}(m^2_{K_1(1400)})\Big] ~, \non \\
  A(D^0\to K^-_1(1270)\pi^+) &=&
 {G_F\over\sqrt{2}}V_{cs}^*V_{ud}\Big[2a_1m_{K_1(1270)}f_\pi(\sin\theta_{K_1}
 V_0^{DK_{1A}}(m_\pi^2)+\cos\theta_{K_1} V_0^{DK_{1B}}(m_\pi^2))\Big] ~, \non \\
 A(D^0\to K^-_1(1400)\pi^+) &=&
 {G_F\over\sqrt{2}}V_{cs}^*V_{ud}\Big[2a_1m_{K_1(1400)}f_\pi(\cos\theta_{K_1}
 V_0^{DK_{1A}}(m_\pi^2)-\sin\theta_{K_1} V_0^{DK_{1B}}(m_\pi^2))\Big] ~, \non\\
 A(D^0\to \ov K^0_1(1270)\pi^0) &=&
 -{G_F\over2}V_{cs}^*V_{ud}
 \Big[2a_2m_{K_1(1270)}f_{K_1(1270)}F_1^{D\pi}(m^2_{K_1(1270)})\Big] ~, \\
 A(D^0\to \ov K^0_1(1400)\pi^0) &=&
 -{G_F\over 2}V_{cs}^*V_{ud}\Big[
 2a_2m_{K_1(1400)}f_{K_1(1400)}F_1^{D\pi}(m^2_{K_1(1400)})\Big] ~, \non \\
 A(D^0\to K^+_1(1270)K^-) &=&
 -{G_F\over \sqrt{2}}V_{cs}^*V_{us}\Big[
 2a_1m_{K_1(1270)}f_{K_1(1270)}F_1^{DK}(m^2_{K_1(1270)})\Big] ~, \non \\
 A(D^0\to K^-_1(1270)K^+) &=&
 {G_F\over \sqrt{2}}V_{cs}^*V_{us}\Big[
 2a_1m_{K_1(1270)}f_{K}(\sin\theta_{K_1} V_0^{DK_{1A}}(m^2_{K})
 +\cos\theta_{K_1} V_0^{DK_{1B}}(m_K^2))\Big] ~,
 \non
 \en
where we have taken into account the $K_{1A}-K_{1B}$ mixing given by Eq.~(\ref{eq:K1mixing}) and neglected the short-distance factorizable $W$-exchange contributions.  Likewise, the $D\to Ka_1(1260)$ and $D\to K b_1(1235)$ decay amplitudes read
 \be
 A(D^+\to \ov K^0a_1^+(1260)) &=&
 -{G_F\over\sqrt{2}}V_{cs}^*V_{ud}
 \left[
 2a_1f_{a_1}m_{a_1}F_1^{DK}(m_{a_1}^2)-2a_2f_Km_{a_1}V_0^{Da_1}(m_K^2)
 \right] ~,
 \non \\
 A(D^0\to K^-a_1^+(1260)) &=& -{G_F\over\sqrt{2}}V_{cs}^*V_{ud}\,
 2a_1f_{a_1}m_{a_1}F_1^{DK}(m_{a_1}^2) ~,
 \non \\
 A(D^0\to \ov K^0a_1^0(1260)) &=&
 {G_F\over 2}V_{cs}^*V_{ud}\,2a_2f_Km_{a_1}V_0^{Da_1}(m_K^2) ~, \non \\
  A(D^0\to \pi^-a_1^+(1260)) &=&
 -{G_F\over \sqrt{2}}V_{cd}^*V_{ud}\,2a_1f_{a_1}m_{a_1}F_1^{D\pi}(m_{a_1}^2) ~,
  \non \\
 A(D^+\to \ov K^0b_1^+(1235)) &=&
 -{G_F\over\sqrt{2}}V_{cs}^*V_{ud}
 \left[
 2a_1f_{b_1}m_{b_1}F_1^{DK}(m_{b_1}^2)-2a_2f_Km_{b_1}V_0^{Db_1}(m_K^2)
 \right] ~,
 \non \\
 A(D^0\to K^-b_1^+(1235)) &=& -{G_F\over\sqrt{2}}V_{cs}^*V_{ud}\,
 2a_1f_{b_1}m_{b_1}F_1^{DK}(m_{b_1}^2) ~,   \non \\
 A(D^0\to \ov K^0b_1^0(1235)) &=&
 {G_F\over 2}V_{cs}^*V_{ud}\,2a_2f_Km_{b_1}V_0^{Db_1}(m_K^2) ~.
 \en

Using the decay constants and form factors presented in Section~\ref{sec:A}, the predicted rates of $D\to \bar K_1\pi, \bar K a_1,\bar K b_1$ decays are listed in Table~\ref{tab:DtoAPtheory}.  To test the validity of the factorization hypothesis, we focus on the $D^+$ decay to $\ov K_1\pi^+$ and $\ov K^0a_1(1260)$ which are free of contamination from weak annihilations.  We see that the predictions are in agreement with experiment for these $D^+$ decays and hence factorization works for $D\to AP$, just as the case of $D\to SP$. The predicted rates for $D^0$ decays are slightly smaller, which implies the importance of the $W$-exchange contribution to $D^0$ decay modes.  The theoretical calculations presented in Table~\ref{tab:DtoAPtheory} are for the $K_1(1270)$-$K_1(1400)$ mixing angle $\theta_{K_1}=50.8^\circ$.  When the other solution $\theta_{K_1}=-44.8^\circ$ is used, we find the predictions $\B(D^+\to \bar K_1^0(1270)\pi^+)=2.9\%$, $\B(D^+\to \bar K_1^0(1400)\pi^+)=2.0\times 10^{-3}$ and $\B(D^0\to K_1^+(1270)\pi^+)=9.1\times 10^{-5}$, all in sharp disagreement with the data.  Historically, it was first pointed out in Ref.~\cite{Cheng:DAP} that a negative mixing angle $\theta_{K_1}$ is ruled out by the data of $D^+\to\bar K_1^0(1270)\pi^+$ and $D^0\to K^-_1(1270)\pi^+$. \footnote{The calculation of \cite{Cheng:DAP} was performed in the ISGW2 model \cite{ISGW2} which has the opposite sign convention to the CLF model.}
It was realized later that the negative $\theta_{K_1}$ solution is also ruled out
by the experimental measurements of $B\to K_1(1270)\gamma$ and
$B\to K_1(1400)\gamma$ \cite{Cheng:Kstargamma}.

It is of interest to notice that the $D^0\to K_1^\pm(1400)K^\mp$ decay is not kinematically allowed, yet a branching fraction comparable to $\B(D^0\to K_1^\pm(1270)K^\mp\to K^\mp K^\pm\pi^-\pi^+)$ has been observed. Since the width of
$K_1(1400)$ is $174\pm13$ MeV, the $D^0\to K_1^\pm(1400)K^\mp$ decay followed by $K_1^\pm(1400)\to K^\pm\pi^+\pi^-$ is certainly allowed.

\subsection{$D\to TP$ \label{sec:facDTP}}

Since the decay constant of tensor meson vanishes, the factorizable amplitude of $D\to TP$ always involves the expression
 \be
\label{eq:XDTP}
 X^{(DT,P)}  &=& \langle P(q)| (V-A)_\mu|0\rangle \langle T(p)| (V-A)^\mu|D(p_D)\rangle \non \\
 &=& if_P\,\vp^*_{\mu\nu}p_D^\mu p_D^\nu
\left[
 k(m_P^2)+b_+(m_P^2)(m_D^2-m_T^2)+b_-(m_P^2)m_P^2
\right] ~,
 \en
where use has been made of Eq.~(\ref{DTff}). The decay rate is given by Eq.~(\ref{eq:rateTP}).  In general, $TP$ final states are suppressed relative to $PP$ states due to less available phase space. More precisely,
 \be
 {\Gamma(D\to TP)\over \Gamma(D\to P_1P_2)}={2\over 3}\,
 {p_T^5\over p_{_P}}\,\left({m_D\over
 m_T}\right)^4\left| {M(D\to TP)\over M(D\to P_1P_2)}\right|^2 ~,
 \en
where we have changed the notation slightly so that $p_T$ denotes the c.~m.\ momentum of the tensor meson and $p_{_P}$ is the c.~m.\ momentum of the pseudoscalar meson $P_1$ or $P_2$ in the charmed meson rest frame. The kinematic factor ${2\over 3}(p_T^5/p_P)(m_D/ m_T)^4$ is typically of order $(1-4)\times 10^{-2}\,{\rm GeV}^{-4}$. An inspection of Table~\ref{tab:DtoTPtheory} indicates that, in the absence of weak annihilation contributions, the Cabibbo-allowed decays $D^+\to \ov K_2^{*0}\pi^+$ and $D^0\to K_2^{*-}\pi^+$ should have the largest decay rates as they proceed through the color-allowed tree diagram $T$. It is easily seen that all other $W$-emission amplitudes in $D\to a_2\ov K$, $D\to f_2\pi$ and $D\to f_2 \ov K$ are suppressed for various reasons. For example, it is suppressed by the vanishing decay constant of the tensor meson, by the small $f_2-f_2'$ mixing angle, by the parameter $a_2$, or by the Cabibbo angle.

From Table~\ref{tab:DtoTPtheory} we see that the predicted branching ratio of $D^+\to \ov K_2^{*0}\pi^+$ is of order $10^{-5}$, which is about two orders of magnitude smaller than experiment. Indeed, the theoretical calculation gives $|T|=3\times 10^{-7}$ in the unit of ${\rm GeV}^{-1}$, which is much smaller than the value of $|T|$ listed in Table~\ref{tab:DTPfit}. As for the decay $D^0\to K_2^{*-}\pi^+$, its rate is similar to that of $D^+\to\ov K^{*0}\pi^+$ but receives an additional $W$-exchange contribution. A fit of this mode to experiment will require $|E|>|T|$, namely, $W$-exchange dominates over the external $W$-emission. The current measurement of $\B(D^+\to\ov K_2^{*0}K^+)$ is problematic as it is Cabibbo-suppressed and yet the measured rate is larger than $\B(D^+\to\ov K_2^{*0}\pi^+)$.

All the predictions shown in Table~\ref{tab:DtoTPtheory} are too small by at least two orders of magnitude, as originally noticed in Ref.~\cite{ChengTP}. In order to resolve the enormous discrepancy between theory and experiment for $TP$ modes, one may consider possible form factor enhancement and finite width effects.  One may compare the $D\to T$ transition form factors calculated in different models: the relativistic light-front quark model (see Table~\ref{tab:FFDtoA}) and the ISGW and ISGW2 quark models (see Table II of Ref.~\cite{ChengTP}). Since the form factors obtained from different models are of the same order, it is very unlikely that they can be enhanced by one order of magnitude to ameliorate the discrepancy. The finite width effect of the tensor resonances will be discussed in the next section.

\section{Finite Width Effects \label{sec:finitewidth}}

Normally we apply the narrow width approximation to extract the two-body branching fraction $\B(D\to MP)$ from 3-body decay data with $M$ standing for an even-parity meson. There are three cases where the narrow width approximation is not valid or justified and the finite width of the resonance has to be taken into account: (i) The decay $D\to MP$ is not kinematically allowed.  For example, $D^+\to \ov K_0^{*0}K^+,\ov K_2^{*0}K^+$ and $D^0\to a_0^+(1450)K^-, K_1^+(1400)K^-$ are forbidden if the scalar resonances are very narrow and on their mass shells. (ii) The resonance width is not negligible.  For example, the widths of $\sigma$ and $\kappa$ are very broad, of order $600-1000$ and $550\pm34$ MeV, respectively \cite{PDG}.  (iii) The strong decay of resonance is marginally allowed or even forbidden kinematically.  For instance, the central values of the $f_0(980)$ and $a_0(980)$ masses are below the threshold for decaying into a charged kaon pair.

In general, the rate of the three-body decay $D\to P_1P_2P$ is given by Ref.~\cite{Chengf01370} \footnote{The case for $D\to AP\to VP_1P\to P_1P_2P_3P$ is more complicated and has been discussed in Ref.~\cite{ChengAP}. The formula for $\Gamma(D\to TP\to P_1P_2P)$ given in Ref.~\cite{ChengTP} was erroneous and it is corrected here.}
\be \label{eq:SP3body}
 \Gamma(D\to SP\to P_1P_2P) &=& {1\over
 2m_D}\int^{(m_D-m_P)^2}_{(m_1+m_2)^2}{dq^2\over 2\pi}\,|M(D\to SP)|^2\,{\lambda^{1/2}(m_D^2,q^2,m_P^2)\over 8\pi m_D^2}  \non \\
 &\times& {1\over (q^2-m_S^2)^2+(\Gamma_{12}(q^2)m_S)^2}\,g^2_{SP_1P_2}
 {\lambda^{1/2}(q^2,m_1^2,m_2^2)\over 8\pi q^2}
 \en
via a scalar resonance, and
\be \label{eq:TP3body}
 \Gamma(D\to TP\to P_1P_2P) &=& {1\over
 \pi}\int^{(m_D-m_P)^2}_{(m_1+m_2)^2}{dq^2\over 2\pi}\,|M(D\to TP)|^2\,{ p(q^2)^5\over 12\pi}\,{m_D^2\over q^2}  \non \\
 &\times& {1\over (q^2-m_T^2)^2+(\Gamma_{12}(q^2)m_T)^2}\,g^2_{TP_1P_2}
 { p'(q^2)^5\over 15\pi q^5}
 \en
 via a tensor resonance, where $\lambda$ is the usual triangular function $\lambda(a,b,c)=a^2+b^2+c^2-2ab-2bc-2ca$, $m_1$ ($m_2$) is the mass of $P_1$ ($P_2$), $p(q^2) = \lambda^{1/2}(m_D^2, q^2,m_P^2)/(2m_D)$, $p'(q^2) = \lambda^{1/2}(q^2,m_1^2,m_2^2)/(2\sqrt{q^2})$, and $g_{MP_1P_2}$ is the strong coupling to be defined below.  The ``running" or ``comoving" width $\Gamma_{12}(q^2)$ is a function of the invariant mass $m_{12}=\sqrt{q^2}$ of the $P_1P_2$ system and has the expression \cite{Pilkuhn}
\be
\Gamma_{12}(q^2) =\left\{
\begin{array}{ll}
\Gamma_T\,{m_T\over m_{12}}\left({p'(q^2)\over
 p'(m_T^2)}\right)^5\,{9+3R^2p'^2(m_T^2)+R^4p'^4(m_T^2)\over 9+
 3R^2p'^2(q^2)+R^4p'^4(q^2)}, & {\rm for}~M=T ~, \\
 \Gamma_S\,{m_S\over m_{12}}{p'(q^2)\over
 p'(m_S^2)} & {\rm for}~ M=S ~.
\end{array}\right.
 \en
 Note that the propagator of the resonance has been assumed to be of the Breit-Wigner form. From the measured widths of the resonances, one can determine their strong couplings
\be
  \Gamma(S\to P_1P_2)=g_{SP_1P_2}^2\,{p_c\over 8\pi m_S^2} ~, \qquad
  \Gamma(T\to P_1P_2)={g_{TP_1P_2}^2 m_T\over 15\pi}\,
  \left({p_c\over m_T}\right)^5 ~.
\en
When the resonance width $\Gamma_M$ is narrow, the expression of the resonant decay rate can be simplified by applying the so-called narrow width approximation
 \be
 {1\over (q^2-m_{M}^2)^2+m_{M}^2\Gamma_{M}^2(q^2)}\approx {\pi\over
 m_{M}\Gamma_{M}}\,\delta(q^2-m_{M}^2) ~.
 \en
It is easily seen that this leads to the factorization relation
Eq.~(\ref{eq:fact}) for the resonant three-body decay.

In the following, we illustrate the finite width effects with a few examples.

\vskip 0.3cm\noindent\underline{$D^+\to \ov K_0^{*0}K^+\to K^+K^-\pi^+$}

With a width of $270\pm80 $ MeV for $K_0^*(1430)$, the decay $D^+\to \ov K_0^{*0}K^+$ followed by $\ov K^{*0}_0\to K^+K^-\pi^+$ is now physically allowed. In this case one should evaluate the two-step process $\Gamma(D^+\to \ov K_0^{*0}K^+\to K^+K^-\pi^+)$ and compare the resonant three-body rate with experiment.  Using Eq.~(\ref{eq:SP3body}) and assuming that the coupling $g_{SP_1P_2}$ is insensitive to the variation in $q^2$ when the resonance is off its mass shell, we obtain
\be
\B(D^+\to \ov K_0^{*0}K^+\to K^+K^-\pi^+)=(1.3^{+0.1}_{-0.3})\times 10^{-4} ~.
\en
This is one order of magnitude smaller than the experimental value, $(1.83\pm0.35)\times 10^{-3}$ (see Table~\ref{tab:DataSP}). Since
\be
A(D^+\to \ov K_0^{*0}K^+) = V_{cs}^*V_{us}T+V_{cd}^*V_{ud}A ~,
\en
the inclusion of $W$-annihilation $A$ will improve the discrepancy between theory and experiment.

\vskip 0.3cm\noindent\underline{$D^+\to\sigma\pi^+$}

Since the width of the $\sigma$ resonance $\Gamma_\sigma=600-1000$ MeV is of the same order of magnitude as its mass, it is important to see its effect on the extraction of the branching fraction $\B(D^+\to\sigma\pi^+)$.  To see this, we shall first define a quantity
\be  \label{eq:eta}
 \eta\equiv {\Gamma(D\to MP\to P_1P_2P)\over \Gamma(D\to MP) \B(M\to P_1P_2)} ~.
 \en
  As $\eta$ goes to 1 in the narrow width approximation, the deviation of $\eta$ from unity gives a measure of violation in the factorization relation (\ref{eq:fact}).  We first compute $\eta$ theoretically; that is, both $\Gamma(D\to MP\to P_1P_2P)$ and $\Gamma(M\to P_1P_2)$ are computed in a model.  The ratio $\eta$ is independent of the form factor for $D\to M$ transition.   The factorization relation Eq.~(\ref{eq:fact}) is then replaced by
 \be \label{eq:facrelation}
 \Gamma(D\to MP\to P_1P_2P)=\eta\,\Gamma(D\to MP)\B(M\to P_1P_2) ~.
 \en
From the experimental input of $\B(D\to MP\to P_1P_2P)$, we can then determine $\Gamma(D\to MP)$.

For $D^+\to\sigma\pi^+\to \pi^+\pi^+\pi^-$ decays, we find $\eta=0.55$ for $\Gamma_\sigma=600$ MeV and $\eta=0.41$ for $\Gamma_\sigma=1000$ MeV, where we have taken $\B(\sigma\to\pi^+\pi^-)={2\over 3}$. This means that the branching fraction $\B(D^+\to\sigma\pi^+)=(2.1\pm0.2)\times 10^{-3}$ listed in Table~\ref{tab:DataSP} should be enhanced by a factor of $1/\eta=1.8\sim 2.4$, depending on the $\sigma$ width.

\vskip 0.3cm\noindent\underline{$D\to TP$}

We have examined the finite-width effect on the $D\to TP$ channels listed in Table~\ref{tab:DtoTPtheory}.  The measured decay widths of various tensor mesons are of order 100 MeV \cite{PDG}.

The singly Cabibbo-suppressed decay $D^+\to \ov K_2^{*0}K^+\to K^+K^-\pi^+$ is physically allowed due to the width $\Gamma_{K_2^{*0}}=109\pm5$ MeV. From Eq.~(\ref{eq:TP3body}) we obtain
\be
\B(D^+\to \ov K_2^{*0}K^+\to K^+K^-\pi^+)=4.3\times 10^{-8} ~.
\en
This is about 3 to 4 orders of magnitude below the experimental result $(1.7^{+1.2}_{-0.8})\times 10^{-4}$ \cite{PDG}.  However, this measurement seems to be problematic as the branching fraction of the Cabibbo-favored mode $D^+\to \ov K_2^{*0}\pi^+\to K^-\pi^+\pi^+$ is of the same order of magnitude, $(2.1\pm0.4)\times 10^{-4}$.

For the other $D\to TP$ decays, we compute the ratio $\eta$ defined in Eq.~(\ref{eq:eta}) and find that $\eta\sim 1.0-1.2$ for most cases except for $D^0\to f_2\ov K^0$ where $\eta=4.0$.  This means that the branching fraction of $D^0\to f_2\ov K^0$ extracted in Table~\ref{tab:TPData} should be reduced by a factor of 4 when the effect of finite width is taken into account\footnote{Our conclusion for finite-width effects on $D\to TP$ differs from that in Ref.~\cite{ChengTP}.}.

\section{Discussions \label{sec:discussions}}

\subsection{The covariant light-front model}

We have relied heavily on the CLF model to obtain the form factors needed in this work. This relativistic quark model preserves the Lorentz covariance in the light-front framework and has been applied successfully to describe various properties of pseudoscalar and vector mesons~\cite{Jaus99}. The analysis of the CLF model has been generalized to even-parity, $P$-wave mesons in \cite{CCH}. Since relativistic effects can manifest in heavy-to-light transitions at large to maximum recoil where the final-state meson becomes highly relativistic, the use of the non-relativistic quark model is probably not suitable here. Therefore, we believe that the CLF approach can provide more accurate behaviors of $B\to M$ transitions at large recoil.  For example, the tensor form factors for $B\to K_1$ transitions at $q^2=0$ derived in the CLF model lead to a prediction for $\B(B\to K_1(1270)\gamma)$ in much better agreement with experiment than any other models (see Table~IV of \cite{Cheng:BAgamma}).

\subsection{Theoretical uncertainties}

Thus far we have not given the error bars to the theoretical results of decay rates. Here we discuss the possible sources of uncertainties in this study. In the CLF model we use the decay constants together with the given constituent quark masses to determine a fundamental parameter $\beta$ in the model for describing the meson wave functions \cite{CCH}. For the decay constants of scalar mesons we use those obtained in \cite{CCY} where QCD sum rules are employed and errors are given explicitly. For axial-vector mesons, the sum rule approach gives $f_{a_1}=238\pm 10$ MeV \cite{YangNP} while the experimental data of $\tau\to K_1\nu$ yield $f_{K_1(1270)}=-(170\pm20)$ MeV and $f_{K_1(1400)}=-(139\pm43)$ MeV [cf. Eqs.~(\ref{eq:fK1abs}) and (\ref{eq:fK1})]. The error bars will be propagated from decay constants to the parameter $\beta$, and then finally to the form factors of interest. Therefore, the uncertainty analysis in the CLF model is quite involved and highly nontrivial. We will leave this task to a future publication.

\subsection{Comparison with other works}

In this work we have performed the study of the nonleptonic decays of charmed mesons in two approaches: flavor-diagram analysis and naive factorization.  In the latter approach, form factors are obtained from the CLF quark model and the weak annihilation diagrams are neglected at the outset. In the literature most of the relevant studies are also based on the factorization approach, differing mainly in the values of the form factors to be used and the treatment of weak annihilation. In Tables~\ref{tab:FFDtoS}-\ref{tab:FFDtoT} we have displayed in parentheses the form factors evaluated in the ISGW2 model \cite{ISGW2}, an improved version of the non-relativitsic quark model by Isgur, Scora, Grinstein, and Wise (ISGW) \cite{ISGW}. We see that the ISGW2 model predicts much smaller $D\to S$ form factors than the CLF model and other models not listed in Table~\ref{tab:FFDtoS}.  In contrast, the form factors $V_0^{Da_1}$, $V_0^{Db_1}$, $k^{Da_2}$, $k^{Df_{2q}}$, $k^{D_s f_{2s}}$ and $k^{D_sK_2^*}$ calculated by the ISGW2 model are much larger than the CLF results.

As stressed before, the factorization hypothesis is best tested in the decays in which weak annihilation contributions are absent or suppressed. For the $SP$ modes, the Cabibbo-allowed decays $D^+\to \ov K_0^{*0}\pi^+$ and $D_s^+\to f_0\pi^+$ belong to this category. It turns out that the former mode is ideal for testing different form factor models. The contribution to $D^+\to \ov K_0^{*0}\pi^+$ from the color-suppressed tree amplitude $C$ was not considered in \cite{Katoch,Fajfer}, presumably due to the smallness of the $K_0^*$ decay constant. Both contributions of color-allowed and color-suppressed amplitudes were taken into account in \cite{Kamal}. However, owing to the destructive interference in $D^+\to \ov K_0^{*0}\pi^+$ and the smallness of $a_2$, the predictions of $\Gamma(D^+\to \ov K_0^{*0}\pi^+)<\Gamma(D^0\to \ov K_0^{*-}\pi^+)$ and the large suppression of $D^0\to \ov K^{*0}\pi^0$ relative to $D^0\to \ov K^{*-}\pi^+$ given in \cite{Kamal} are not borne out by experiment. In the CLF model, the relative sign of the factorizable amplitudes $X^{(DS,P)}$ and $X^{(DP,S)}$ are fixed to be negative [see Eq.~(\ref{eq:XSP})]. As a consequence, we conclude that, based on the CLF quark model, the interference in the Cabibbo-allowed $D^+\to SP$ decays must be constructive, contrary to the case of $D\to PP,VP$.

Recently, the decays $D\to K_1\pi$ have been considered in \cite{Sharma,Khosravi} with $D\to K_1$ from factors evaluated in the framework of the ISGW model and QCD sum rules, resepctively. While the predicted rates are similar to ours in most cases, the branching fraction of $D^+\to \ov K^0_1(1270)\pi^+$ was found to be $(5.85\pm0.37)\%$ for $\theta_{K_1}=-37^\circ$ and $(3.18\pm 0.25)\%$ for $\theta_{K_1}=-58^\circ$ in \cite{Khosravi}. All of them are too large compared to the experimental limit $7\times 10^{-3}$ \cite{PDG}.  In our study we did include the color-suppressed tree amplitude [see Eq.~(\ref{eq:K1pi})], which was considered in \cite{Sharma} but neglected in \cite{Khosravi}. Since the decay constant of $K_1(1270)$ is negative [Eq.~(\ref{eq:fK1})], it is clear from Eq.~(\ref{eq:K1pi}) that the $a_2$ term contributes destructively to $D^+\to \ov K^0_1(1270)\pi^+$. Consequently, our prediction $\B(D^+\to \ov K^0_1(1270)\pi^+)=4.7\times 10^{-3}$ (Table~\ref{tab:DtoAPtheory}) is consistent with experiment. \footnote{$\B(D^+\to \ov K^0_1(1270)\pi^+)$ is predicted to be $3.8\times 10^{-3}$, $1.52\%$ and $3.21\%$ in \cite{Sharma} for $\theta_{K_1}=33^\circ,45^\circ$ and $57^\circ$, respectively. However, as noticed in Sec.~III.B, the mixing angle $\theta_{K_1}$ has to be negative in the convention of QCD sum rules or the ISGW model. Therefore, the mixing angle chosen by \cite{Sharma} does not have a correct sign. Also the relative sign between $T$ and $C$ amplitudes given there is erroneous.}

The above few examples indicate that the CLF model takes care of the relative signs of decay constants, magnitudes of form factors, and various hadronic matrix elements correctly.

\subsection{Comparison with $B$ decays}

Charmful decays $B\to D^{**}P$ with $D^{**}=D_0^*,D_1,D'_1,D_2^*$ denoting even-parity charmed mesons and charmless decays $B\to MP$ ($M$ being light even-parity mesons) have been extensively studied both experimentally \cite{PDG} and theoretically \cite{Cheng:BtoDpwave,CCY,Cheng:BtoSV,Cheng:BtoAP}.  It is instructive to compare the present work with the $B$ decays.

While factorization works well for Cabibbo-allowed $D^+ \to SP, AP$ decays, predictions are typically about one order of magnitude smaller than experiment for the other decay modes, conceivably due to the negligence of weak annihilation contributions arising from final-state interactions.  It is pointed out in \cite{CCY} that one needs a sizable penguin weak annihilation amplitude in order to account for the data of $B\to K^*_0\pi$ in the theoretical approach such as QCD factorization.

Taking the cue from the constructive interference in the Cabibbo-allowed decay $D^+\to \ov K_0^{*0}\pi^+$ and noting that the phase of $a_2/a_1$ in $B\to D\pi$ decays lies in the first quadrant, one may be tempted to claim that the $B^-\to D^{*0}_0\pi^+$ decay should have a rate smaller than the $B^0\to D^{*-}_0\pi^+$ decay owing to a destructive interference in the former. This conjecture is supported in both the CLF model and heavy quark symmetry \cite{Cheng:BtoDpwave}. The experimental observation that the production of
broad $D^{**}$ states in charged $B$ decays is more than a factor
of five larger than that produced in neutral $B$ decays (see Tables~V and VI in \cite{Cheng:BtoDpwave}) is thus astonishing. This enigma as pointed out several years ago in \cite{Cheng:BtoDpwave} still remains unresolved.

\section{Conclusions \label{sec:conclusions}}

In this paper, we have studied the charmed meson decays into final states containing one pseudoscalar meson and one even-parity meson, the data of which are inferred from detailed Dalitz analyses of three-body decays and the finite width effects.  The non-perturbative flavor diagram approach and the factorization calculation are undertaken to analyze these decay processes.  In the diagrammatic framework, we have extracted the sizes and relative strong phases of various flavor diagrams in a least model-dependent way, based on current experimental measurements of decay rates.  In the factorization approach, we have neglected weak annihilation diagrams ($E$ and $A$) while making our predictions.  Besides, we use the form factors evaluated in the covariant light-front model to compute the $D$ decay branching fractions.

In the $D \to SP$ decays with $S$ being in the non-strange nonet, our fits present several robust features against the 2- and 4-quark pictures.  First, as in the $PP$ and $VP$ decays, the weak annihilation diagrams are non-negligible.  However, it is a theoretically challenging puzzle that the $A$ amplitude here is the largest one.  Secondly, the relative strong phases $\delta_A$ and $\delta_E$ are preferably around $30^\circ$ and $160^\circ$, respectively.  Finally, the data do not prefer either the 2-quark scheme or the 4-quark scheme.  This conclusion can be best seen from a comparison between the $D^{0,+} \to f_0 \pi$ and $\sigma\pi$ decays, and also revealed in the $\chi^2_{\rm min}$ values in Table~\ref{tab:DSPfit}.  We note that due to a paucity of measured modes, the diagrammatic approach is ineffective for the $D \to K_0^* P$ decays or the $D \to AP$ decays.  Theoretical calculations based on factorization have been applied to both $D \to SP$ and $AP$ transitions.  The factorization hypothesis seems to work pretty well for those Cabibbo-favored modes that involve only the color-allowed tree ($T$) and color-suppressed tree ($C$) amplitudes.  The factorization calculations for the other decay modes are typically about one order of magnitude smaller than experiment, conceivably due to the negligence of weak annihilation contributions.

Contrary to the Cabibbo-allowed $D^+\to PP,VP$ decays where $T$ and $C$ amplitudes interfere destructively, the color-allowed and color-suppressed tree amplitudes in the Cabibbo-allowed decays $D^+\to \ov K^{*0}\pi^+,\ov K_1^0(1400)\pi^+$ contribute constructively.  This explains why their branching fractions  are large, of ${\cal O}(2\%)$.

The $D \to TP$ measurements poise the biggest problem for theory.  Even though the magnitudes and phases of some amplitudes can be extracted from data, as given in Table~\ref{tab:DTPfit}, quite opposite conclusions are reached when different sets of data [Fit (A) versus Fit (B)] are used.  This could be caused by the $D^+ \to \ov K_2^{*0} \pi^+$ and $D^+ \to K_2^{*0} \pi^+$ decays, as explained in the text.  Predicted branching fractions based on factorization are at least two orders of magnitude smaller than data, even when the decays are free of weak annihilation contributions.  We cannot find possible sources of rate enhancement.

We also examine the finite width effects for decays that are kinematically forbidden if the width of the even-parity meson is not taken into account.  We find that the branching fraction of $D^+\to\sigma\pi^+$  extracted from three-body decays is enhanced by a factor of 2, whereas $\B(D^0\to f_2(1270)\ov K^0)$ is reduced by a factor of 4 by finite width effects.

Our study shows that some of the above-mentioned puzzles call for measurements of yet observed decay modes as well as more precise determination in the decay rates.  The other puzzles, particularly in the $TP$ modes, demand better understanding of the underlying dynamics.

\section*{Acknowledgments}
H.-Y.~C. wishes to thank the hospitality of the Physics Department, Brookhaven National Laboratory.  This research was supported in part by the National Science Council of Taiwan, R.~O.~C.\ under Grant Nos.~NSC~97-2112-M-008-002-MY3, NSC~97-2112-M-001-004-MY3 and in part by the NCTS.

\end{document}